\documentclass[aps,prd,reprint,amsmath,amssymb]{revtex4-2}
\usepackage{graphicx,tikz,multirow,pgfplots,array,booktabs}
\pgfplotsset{compat=1.3}
\usepackage{newtxtext,newtxmath}
\definecolor{citecolor}{RGB}{45,47,146}
\usepackage[colorlinks,citecolor=citecolor,anchorcolor=red,menucolor=red, linkcolor=citecolor,filecolor=red,runcolor=red,urlcolor=citecolor,frenchlinks=red]{hyperref}

\begin{document}
    \let\oldcite=\cite
    \renewcommand{\cite}[1]{\textcolor{citecolor}{\oldcite{#1}}}
    \renewcommand{\eqref}[1]{\textcolor{citecolor}{(\ref{#1})}}
    
    \title{Triply heavy baryon spectroscopy revisited
    }
    \author{Hao Zhou$^{1,2,4}$}
    \author{Si-Qiang Luo$^{1,2,4}$}\email{luosq15@lzu.edu.cn}
    \author{Xiang Liu$^{1,2,3,4}$}\email{xiangliu@lzu.edu.cn}
    \affiliation{
        $^1$School of Physical Science and Technology, Lanzhou University, Lanzhou 730000, China\\
        $^2$Lanzhou Center for Theoretical Physics,
        Key Laboratory of Theoretical Physics of Gansu Province,
        Key Laboratory of Quantum Theory and Applications of MoE,
        Gansu Provincial Research Center for Basic Disciplines of Quantum Physics, Lanzhou University, Lanzhou 730000, China\\
        $^3$MoE Frontiers Science Center for Rare Isotopes, Lanzhou University, Lanzhou 730000, China\\
        $^4$Research Center for Hadron and CSR Physics, Lanzhou University $\&$ Institute of Modern Physics of CAS, Lanzhou 730000, China}
	
    \begin{abstract}
		We present a comprehensive study of triply heavy baryons ($\Omega_{ccc}$, $\Omega_{bbb}$, $\Omega_{bcc}$, and $\Omega_{bbc}$) within the nonrelativistic quark model, employing the Gaussian expansion method to calculate mass spectra up to $D$-wave states. Our analysis represents the most complete treatment to date for this model, incorporating full angular momentum mixing effects. While our predictions for low-lying states agree well with lattice quantum chromodynamics (QCD) results, we find systematically lower masses for excited states compared to lattice calculations. Using the obtained wave functions, we estimate radiative decay widths up to $1D$ states, revealing significant differences from previous theoretical work. Additionally, we identify and resolve several misconceptions in prior treatments of triply heavy baryon spectroscopy, particularly symmetry constraint and wave function construction in three-quark systems. These results provide crucial information for future experimental searches and theoretical investigations of triply heavy baryon systems. 
    \end{abstract}

    \maketitle
    \section{INTRODUCTION}

Hadron spectroscopy is an effective approach to deepen our understanding of the non-perturbative behavior of the strong interaction, a challenging task in contemporary particle physics. With the accumulation of experimental data over the past two decades, numerous new hadronic states have been observed \cite{Liu:2013waa,Hosaka:2016pey,Chen:2016qju,Richard:2016eis,Olsen:2017bmm,Guo:2017jvc,Liu:2019zoy,Brambilla:2019esw,Meng:2022ozq,Chen:2022asf}. The study of hadron spectroscopy is entering a new period, in which the characterization of conventional hadrons and the exploration of exotic hadronic states are becoming central research focuses.

Although the existence of triply heavy baryons is theoretically inevitable, the hunt for them is still underway, since they require the simultaneous production and binding of three heavy quarks, making their formation exceedingly rare. In 2017, the observation of the double-charm baryon $\Xi_{cc}^{++}(3621)$ by the LHCb Collaboration \cite{LHCb:2018pcs} raised expectations for new progress in the search for triply charmed baryons, especially with the upcoming high-luminosity upgrade of the LHC.

In fact, initial research on the spectroscopy of triply heavy baryons dates back to the bag model in 1980s \cite{Hasenfratz:1980ka,Bernotas:2008bu}. Later, the mass spectra of triply heavy baryons have been extensively studied using various theoretical frameworks, including relativistic or nonrelativistic quark model \cite{Silvestre-Brac:1996myf,Vijande:2004at,Migura:2006ep,Jia:2006gw,Martynenko:2007je,Roberts:2007ni,Patel:2008mv,Flynn:2011gf,Shah:2017jkr,Weng:2018mmf,Qin:2018dqp,Yang:2019lsg,Liu:2019vtx,Faustov:2021qqf,Ortiz-Pacheco:2023kjn,Yu:2025gdg}, lattice QCD \cite{Dhindsa:2024erk,Padmanath:2013zfa,Mathur:2018epb,Brown:2014ena,Burch:2015pka,Meinel:2010pw,Meinel:2012qz,Alexandrou:2023dlu,Li:2022vbc,Lyu:2021qsh,Bahtiyar:2020uuj,Alexandrou:2017xwd,Chen:2017kxr,Can:2015exa,Alexandrou:2014sha,PACS-CS:2013vie,Durr:2012dw,Alexandrou:2012xk,Briceno:2012wt,Chiu:2005zc}, potential nonrelativistic QCD (pNRQCD) \cite{Brambilla:2005yk,Llanes-Estrada:2011gwu}, QCD sum rules \cite{Zhang:2009re,Wang:2011ae,Aliev:2012tt,Aliev:2014lxa,Azizi:2014jxa,Wang:2020avt,Najjar:2025dzl}, Regge phenomenology \cite{Guo:2008he,Wei:2015gsa,Wei:2016jyk} and others \cite{Serafin:2018aih,Yin:2019bxe,Gutierrez-Guerrero:2019uwa,Gomez-Rocha:2023jfr}. These continued investigations also underscore the persistent attention given to triply heavy baryons in hadron physics.

Among these theoretical studies, Refs.~\cite{Vijande:2004at,Migura:2006ep,Jia:2006gw,Martynenko:2007je,Roberts:2007ni,Patel:2008mv,Flynn:2011gf,Weng:2018mmf,Qin:2018dqp,Ortiz-Pacheco:2023kjn} focus exclusively on the low-lying states of triply heavy baryons, whereas Refs.~\cite{Silvestre-Brac:1996myf,Shah:2017jkr,Yang:2019lsg,Liu:2019vtx,Faustov:2021qqf,Yu:2025gdg} provide systematic calculations of their mass spectra. Nevertheless, these comprehensive investigations remain incomplete. 
For example, the interaction potentials used in Refs.~\cite{Silvestre-Brac:1996myf,Yang:2019lsg} include only central terms, making them unable to describe the fine structure of triply heavy baryons. Furthermore, the calculation in Ref.~\cite{Liu:2019vtx} neglected spin $S=1/2$ and $S=3/2$ state mixing---an omission that significantly impacts radiative decay computations---and was limited to the mass spectra of $\Omega_{ccc}$ and $\Omega_{bbb}$ baryons. Additionally, Ref.~\cite{Faustov:2021qqf} employs a quark-diquark model, whereas Ref.~\cite{Shah:2017jkr} uses the hyperspherical approximation. Finally, Ref.~\cite{Yu:2025gdg} utilized only a single incomplete set of Jacobi coordinates and omitted decay studies.
Given these limitations, a renewed investigation into the spectroscopy of triply heavy baryons---particularly higher-precision hadron spectroscopy---is warranted.

Over the past several years, we have successfully established theoretical descriptions of the mass spectra and decay properties of singly heavy baryons \cite{Luo:2023sra,Luo:2023sne,Peng:2024pyl} and $\Omega$ hyperons \cite{Luo:2025cqs} within the nonrelativistic quark model framework, additionally predicting new hadronic state \cite{Luo:2021dvj}. These studies demonstrate that the effective potential employed in these works accurately reproduces hadronic spectroscopy. This validated formalism will now be extended to the present study of triply heavy baryons.

Triply heavy baryons constitute few-body systems. To address such systems with high precision, we employ the Gaussian Expansion Method (GEM) \cite{Kamimura:1988zz,Hiyama:2003cu,Hiyama:2012sma}. In this approach, the wave function of a few-body system is expanded into a set of Gaussian basis functions, transforming the solution of the stationary Schr\"odinger equation into a generalized eigenvalue problem. The Gaussian basis functions incorporate different angular momentum excitations, and the relative contributions of these excitations vary across different states and systems. Notably, a recent study \cite{Yu:2025gdg} observed that the lowest state of the $\Omega_{bcc}$ baryon is dominated by the $\lambda$-mode, whereas the $\Omega_{bbc}$ baryon exhibits a dominant $\rho$-mode orbital excitation. In fact, this phenomenon was found and explained by Ref.~\cite{Yoshida:2015tia} as early as 2015. We will investigate this phenomenon further from a symmetry perspective. Additionally, we will comprehensively account for angular momentum mixing effects—such as between total spin states ($S=1/2$ and $S=3/2$) and $S$-$D$ mixing—to improve the description of the system's fine structure. Finally, we will address minor inaccuracies in earlier treatments of three-body systems composed of identical particles \cite{Yang:2019lsg,Liu:2019vtx,Yu:2025gdg}.

Owing to their high strong decay thresholds, triply heavy baryons are expected to decay predominantly via radiative transitions. However, only a limited number of studies \cite{Liu:2019vtx,Ortiz-Pacheco:2023kjn} have systematically calculated their radiative decay widths. Here, we present a systematic calculation of these decays. The nonrelativistic radiative decay framework, originally developed in Refs. \cite{Deng:2015bva,Deng:2016stx,Deng:2016ktl}, successfully described the radiative decays of charmonium and bottomonium by employing a multipole expansion of electromagnetic interactions (E$l$ and M$l$ transitions). In our work, we calculate the full transition amplitude directly, bypassing the multipole expansion, thereby achieving greater accuracy within the nonrelativistic approximation.

We note that Ref. \cite{Liu:2019vtx} neglected the mixing between total spin states ($S=1/2$ and $S=3/2$) in its mass spectrum calculations. This omission is significant because the total spin $S$ of the initial and final baryons determines whether an electric transition is allowed, directly influencing the radiative decay width. Furthermore, the wave function of a baryon composed of three identical quarks must adhere to fermionic symmetry, leading to distinct behavior compared to other baryon types. We will analyze how these effects shape the radiative decay properties of triply heavy baryons.

This paper is organized as follows. Following the introduction, we provide a brief review of the theoretical framework used in this work (Sec. \ref{sec2}). In Sec. \ref{sec3}, we present numerical results for the mass spectra and radiative decays of the studied triply heavy baryons. Finally, we conclude with a summary.

    \section{Theoretical framework}\label{sec2}
    
    This section outlines the model and methodology employed in our study. First, we derive the Hamiltonian and determine the model parameters for triply heavy baryons within the nonrelativistic quark model. Next, we construct the corresponding wave functions. We then numerically solve the stationary Schr\"odinger equation using the GEM. Finally, we utilize the obtained wave functions to calculate radiative decay widths of triply heavy baryons.

\subsection{Effective Hamiltonian}

In the conventional quark model framework, triply heavy baryons are composed of three heavy quarks. Following previous works \cite{Luo:2023sra,Luo:2023sne,Peng:2024pyl,Luo:2025cqs,Luo:2021dvj}, we adopt the same Hamiltonian, which is derived from the nonrelativistic limit of the formalism presented in Refs. \cite{Capstick:1986ter,Godfrey:1985xj}. The Hamiltonian is taken as the form:
    \begin{equation}
        \begin{aligned}\label{eq:Hamiltonian}
            H=&\sum_{i}\left(m_i+\frac{p_i^2}{2m_i}\right)\\
            &+\sum_{i<j}\left(V_{ij}^{\text{conf}}+V_{ij}^{\text{hyp}}+V_{ij}^{\text{so(cm)}}+V_{ij}^{\text{so(tp)}}\right).
        \end{aligned}
    \end{equation}
    Here, the potential energy component of the Hamiltonian comprises four distinct terms: (1) the confinement potential $V_{ij}^{\text{conf}}$, which governs quark confinement; (2) the hyperfine potential $V_{ij}^{\text{hyp}}$, responsible for inducing $S$-$D$ wave mixing; (3) the color-magnetic potential $V_{ij}^{\text{so(cm)}}$, describing spin-orbit coupling effects; and (4) the Thomas precession potential $V_{ij}^{\text{so(tp)}}$, arising from relativistic dynamical corrections. These potentials take the following explicit forms:
    \begin{align}
        &V_{ij}^{\text{conf}}=-\frac{2}{3}\frac{\alpha_{s}}{r_{ij}}+\frac{b}{2}r_{ij}+\frac{1}{2}C,\\
        &V_{ij}^{\text{hyp}}=\frac{2\alpha_{s}}{3m_{i}m_{j}}\left[\frac{8\pi}{3}\tilde{\delta}(r_{ij})\boldsymbol{S}_{i}\boldsymbol{\cdot}\boldsymbol{S}_{j}+\frac{1}{r_{ij}^3}S(\boldsymbol{r}_{ij},\boldsymbol{S}_{i},\boldsymbol{S}_{j})\right],\\
        &\begin{aligned}
            V_{ij}^{\text{so(cm)}}=&\frac{2\alpha_{s}}{3r_{ij}^{3}}\left(\frac{\boldsymbol{r}_{ij}\times\boldsymbol{p}_{i}\boldsymbol{\cdot}\boldsymbol{S}_{i}}{m_{i}^{2}}-\frac{\boldsymbol{r}_{ij}\times\boldsymbol{p}_j\boldsymbol{\cdot}\boldsymbol{S}_{j}}{m_{j}^2}\right.\\
            &\left.-\frac{\boldsymbol{r}_{ij}\times\boldsymbol{p}_j\boldsymbol{\cdot}\boldsymbol{S}_i-\boldsymbol{r}_{ij}\times\boldsymbol{p}_i\boldsymbol{\cdot}\boldsymbol{S}_j}{m_im_j}\right),
        \end{aligned}\\
        &V_{ij}^{\text{so(tp)}}=-\frac{1}{2r_{ij}}\frac{\partial V_{ij}^{\text{conf}}}{\partial r_{ij}}\left(\frac{\boldsymbol{r}_{ij}\times\boldsymbol{p}_{i}\boldsymbol{\cdot}\boldsymbol{S}_{i}}{m_{i}^{2}}-\frac{\boldsymbol{r}_{ij}\times\boldsymbol{p}_{j}\boldsymbol{\cdot}\boldsymbol{S}_{j}}{m_{j}^{2}}\right),
    \end{align}
    \begin{equation}
        \tilde{\delta}(r_{ij})=\frac{\sigma^{3}}{\pi^{3/2}}\mathrm{e}^{-\sigma^2r_{ij}^2},\ S(\boldsymbol{r}_{ij},\boldsymbol{S}_{i},\boldsymbol{S}_j)=\frac{3\boldsymbol{S}_i\boldsymbol{\cdot}\boldsymbol{r}_{ij}\boldsymbol{S}_j\boldsymbol{\cdot}\boldsymbol{r}_{ij}}{r_{ij}^{2}}-\boldsymbol{S}_{i}\boldsymbol{\cdot}\boldsymbol{S}_{j},
    \end{equation}
    where $m_i$, $\boldsymbol{p}_i$, and $\boldsymbol{S}_i$ represent the mass, center-of-mass momentum, and spin of the $i$-th quark respectively, while $\boldsymbol{r}_{ij} = \boldsymbol{r}_i - \boldsymbol{r}_j$ denotes the relative position vector between the $i$-th and $j$-th quarks.

The parameters of the model include: $\alpha_s$ (one-gluon-exchange coupling constant), $b$ (linear confinement strength), $C$ (renormalized mass constant), and $\sigma$ (smearing parameter). The parameter determination was performed as follows: the charm quark mass $m_c$ and first row parameters are obtained by fitting the charmonium spectrum; the bottom quark mass $m_b$ and last row parameters are determined from bottomonium spectrum fits; while the middle row parameters are derived through $B_c$ meson mass spectrum fitting.

We employ the $\chi^2$ statistic to quantitatively assess parameter optimization, defined as:
\begin{equation}
\chi^2 = \sum_{i=1}^n \left( \frac{M^{\rm Exp.}_i - M^{\rm The.}_i}{M^{\rm Err.}_i} \right)^2,
\end{equation}
where $M^{\rm Exp.}$, $M^{\rm The.}$, and $M^{\rm Err.}$ denote the experimental masses, theoretical predictions, and their uncertainties, respectively. In practice, parameters are adjusted to minimize $\chi^2$. Since the uncertainties ($M^{\rm Err.}$) for some states are very small, they disproportionately influence the fitting weights, resulting in large contributions to $\chi^2$. Nevertheless, most theoretical predictions agree well with experimental masses. The experimental-theoretical mass comparisons are shown in Table~\ref{tab:comparisons}. And the corresponding parameters listed in Table~\ref{tab:parameters}, with separate parameter sets used for $cc$, $cb$, and $bb$ interactions.

\begin{table}
\centering
\caption{The comparisons of experimental and theoretical masses in $c\bar{c}$, $b\bar{b}$, and $b\bar{c}$ systems. The $M^{\rm Exp.}$, $M^{\rm The.}$, and $M^{\rm Err.}$ are experimental results, theoretical calculations, and uncertainties of the masses, respectively. We also present $\chi^2/n$ here, where $n$ is the number of the particles.}
\label{tab:comparisons}
\renewcommand\arraystretch{1.5}
\begin{tabular*}{86mm}{@{\extracolsep{\fill}}clcc}
\toprule[1.00pt]
\toprule[1.00pt]
States                      &$M^{\rm Exp.}$ (MeV)~\cite{ParticleDataGroup:2024} &$M^{\rm The.}$ (MeV) &$M^{\rm Err.}$ (MeV)~\cite{ParticleDataGroup:2024}\\
\midrule[0.75pt]
$\eta_c(1S)$                &2984.1&2988.9&0.4\\
$\eta_c(2S)$                &3637.7&3642.7&0.9\\
$J/\psi(1S)$                &3096.900&3096.9&0.006\\
$J/\psi(2S)$                &3686.097&3686.1&0.011\\
$h_{c}(1P)$                 &3525.37&3514.4&0.14\\
$\chi_{c0}(1P)$             &3414.71&3439.4&0.30\\
$\chi_{c1}(1P)$             &3510.67&3509.9&0.05\\
$\chi_{c2}(1P)$             &3556.17&3556.1&0.07\\
\multicolumn{4}{c}{$\chi^2/n=1686.4$}\\
\midrule[0.75pt]
$\eta_b(1S)$                &9398.7&9405.1&2.0\\
$\Upsilon(1S)$              &9460.40&9460.6&0.10\\
$\Upsilon(2S)$              &10023.4&10006.6&0.5\\
$h_{b}(1P)$                 &9899.3&9886.3&0.8\\
$h_{b}(2P)$                 &10259.8&10251.9&1.2\\
$\chi_{b0}(1P)$             &9859.44&9838.0&$\pm0.42\pm0.31$\\
$\chi_{b0}(2P)$             &10232.5&10211.4&$\pm0.4\pm0.5$\\
$\chi_{b1}(1P)$             &9892.78&9883.7&$\pm0.26\pm0.31$\\
$\chi_{b1}(2P)$             &10255.46&10252.3&$\pm0.22\pm0.50$\\
$\chi_{b2}(1P)$             &9912.21&9907.1&$\pm0.26\pm0.31$\\
$\chi_{b1}(2P)$             &10268.65&10272.5&$\pm0.22\pm0.50$\\
\multicolumn{4}{c}{$\chi^2/n=453.1$}\\
\midrule[0.75pt]
$B_c(1S)$                   &6274.47&6274.5&0.32\\
$B_c(2S)$                   &6871.2&6871.2&1.0\\
\multicolumn{4}{c}{$\chi^2/n=0.0407$}\\
\bottomrule[1.00pt]
\bottomrule[1.00pt]
\end{tabular*}
\end{table}

    \begin{table}[htbp]
        \caption{The parameters involved in the adopted potential model. \label{tab:parameters}}
        \begin{ruledtabular}
            \begin{tabular}{ccccc}
                &$\alpha_s$&$b (\mathrm{GeV}^2)$&$C (\mathrm{GeV})$&$\sigma (\mathrm{GeV})$\\
                \hline
                $cc$&0.470&0.165&$-0.409$&1.220\\
                $cb$&0.362&0.189&$-0.555$&1.586\\
                $bb$&0.333&0.203&$-0.603$&1.908\\
                \hline
                \multicolumn{5}{c}{$m_c=1.649\,\mathrm{GeV}\quad m_b=5.036\,\mathrm{GeV}$}\\
            \end{tabular}
        \end{ruledtabular}
    \end{table}
	
    \subsection{Wave function}
    \begin{figure}[htbp]
        \begin{tikzpicture}[semithick]
            \fill(-1,0)node[below]{$m_1$}circle[radius=1.5pt];
            \fill(1,0)node[below]{$m_2$}circle[radius=1.5pt];
            \fill(0,1.732)node[above]{$m_3$}circle[radius=1.5pt];
            \draw[-latex](-1,0)--node[above left=-1.5pt]{$\boldsymbol{r}_2$}(0,1.732);
            \draw[-latex](-0.6,0.693)--node[above]{$\boldsymbol{R}_2$}(1,0);
            \node at(0,-0.7){$c=2$};
            \begin{scope}[xshift=-3cm]
                \fill(-1,0)node[below]{$m_1$}circle[radius=1.5pt];
                \fill(1,0)node[below]{$m_2$}circle[radius=1.5pt];
                \fill(0,1.732)node[above]{$m_3$}circle[radius=1.5pt];
                \draw[-latex](0,1.732)--node[above right=-1.5pt]{$\boldsymbol{r}_1$}(1,0);
                \draw[-latex](0.6,0.693)--node[above]{$\boldsymbol{R}_1$}(-1,0);
                \node at(0,-0.7){$c=1$};
            \end{scope}
            \begin{scope}[xshift=3cm]
                \fill(-1,0)node[below]{$m_1$}circle[radius=1.5pt];
                \fill(1,0)node[below]{$m_2$}circle[radius=1.5pt];
                \fill(0,1.732)node[above]{$m_3$}circle[radius=1.5pt];
                \draw[-latex](1,0)--node[below]{$\boldsymbol{r}_3$}(-1,0);
                \draw[-latex](0,0)--node[right]{$\boldsymbol{R}_3$}(0,1.732);
                \node at(0,-0.7){$c=3$};
            \end{scope}
        \end{tikzpicture}
        \caption{Three Jacobian coordinates of three-body system.\label{fig:Jacobian coordinates}}
    \end{figure}
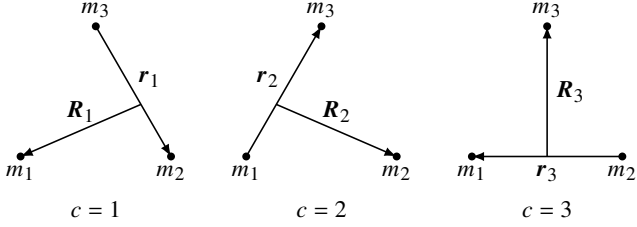

    To determine the mass spectra of triply heavy baryons, we should solve the stationary Schr\"odinger equation corresponding to the Hamiltonian in Eq.~\eqref{eq:Hamiltonian}. Prior to numerical solution, we can identify operators that commute with the Hamiltonian to construct a complete set of commuting observables. This approach provides preliminary insights into the wave function's structure.

For the angular momentum components, we naturally select operators corresponding to $L$-$S$ coupling. The final results are, in principle, independent of the chosen coupling representation. The operator set is
    \begin{equation}\label{eq:operators}
        \{l_c^2, L_c^2, L^2, S_1^2, S_2^2, S_3^2, S_{ij}^2, S^2, J^2, J_z\},
    \end{equation}
    where $\boldsymbol{l}_c$ and $\boldsymbol{L}_c$ represent the orbital angular momenta associated with the $c$-th Jacobi coordinates $\boldsymbol{r}_c$ and $\boldsymbol{R}_c$, respectively, while $\boldsymbol{S}_1$, $\boldsymbol{S}_2$, and $\boldsymbol{S}_3$ denote the spins of the constituent quarks $m_1$, $m_2$, and $m_3$, as illustrated in Fig.~\ref{fig:Jacobian coordinates}. The composite angular momenta are defined as:
\begin{itemize}
    \item $\boldsymbol{L} = \boldsymbol{l}_c + \boldsymbol{L}_c$ (total orbital angular momentum),
    \item $\boldsymbol{S}_{ij} = \boldsymbol{S}_i + \boldsymbol{S}_j$ (pair spin in $\boldsymbol{r}_c$-degree-of-freedom),
    \item $\boldsymbol{S} = \boldsymbol{S}_1 + \boldsymbol{S}_2 + \boldsymbol{S}_3$ (total spin),
    \item $\boldsymbol{J} = \boldsymbol{L} + \boldsymbol{S}$ (total angular momentum).
\end{itemize}

We first examine the commutation relations between these operators (Eq.~\eqref{eq:operators}) and the Hamiltonian. Notably, $l_c^2$ and $L_c^2$ fail to commute with the central potential $V_{ij}(r_{ij})$, requiring the system's wave function to be expressed as a superposition of states with different $l_c$ and $L_c$ quantum numbers. However, $\boldsymbol{L}$ does commute with both $V_{ij}(r_{ij})$ and the dominant spin-spin interaction terms $\boldsymbol{S}_i \cdot \boldsymbol{S}_j$, making $L$ a reasonably good quantum number. Our subsequent calculations will demonstrate that the $S$-$D$ mixing effect is indeed negligible.

For any given $L$, there exist infinitely many possible couplings between $l_c$ and $L_c$. Following established approaches \cite{Kamimura:1988zz}, we employ an expansion using Jacobian coordinates of three rearrangement channels $(c=1\sim3)$ while restricting $l_c$ and $L_c$ to relatively small values. As demonstrated in previous studies \cite{Hiyama:2003cu,Hiyama:2012sma}, this treatment ensures rapid convergence of mass and binding energy calculations due to the inherent advantages of three-channel Jacobi coordinates.

The operator $S_{ij}^2$ commutes exclusively with the contact term $\boldsymbol{S}_i \cdot \boldsymbol{S}_j$ and the tensor term $S(\boldsymbol{r}_{ij}, \boldsymbol{S}_i, \boldsymbol{S}_j)$, whereas $S^2$ only commutes with $\boldsymbol{S}_i \cdot \boldsymbol{S}_j$---a notable departure from two-body systems. Consequently, the non-commuting terms with $S_{ij}^2$ and $S^2$ necessitate the wave function to be a superposition of states with different $s_{ij}$ and $S$ quantum numbers. The total angular momentum $\boldsymbol{J}$, being fully conserved (i.e., commuting with the Hamiltonian), serves as the primary quantum number for state classification.
    
Finally, we discuss a special case where a matrix element vanishes due to symmetry considerations. This case is helpful for understanding the mixing of different modes. If $V$ remains unchanged under the transformation $\boldsymbol{r}_c\to-\boldsymbol{r}_c$ or $\boldsymbol{R}_c\to-\boldsymbol{R}_c$, then the matrix element
\begin{equation}
    \langle l_c'L_c'L'M'|V|l_cL_cLM\rangle=(-1)^{l_c'+l_c}\langle l_c'L_c'L'M'|V|l_cL_cLM\rangle
\end{equation}
or
\begin{equation}
    \langle l_c'L_c'L'M'|V|l_cL_cLM\rangle=(-1)^{L_c'+L_c}\langle l_c'L_c'L'M'|V|l_cL_cLM\rangle.
\end{equation}
This shows that the matrix element of $V$ is not zero only when $l_c'+l_c=\text{even}$ or $L_c'+L_c=\text{even}$, so $V$ can only lead to the mixing of bases of $l_c'+l_c=\text{even}$ or $L_c'+L_c=\text{even}$. The confinement and hyperfine potential of the $\Omega_{bcc}$ and $\Omega_{bbc}$ baryons precisely satisfy transformation invariance, so the mixing of different modes of most of their states is relatively small. But the (0,2,2)-mode and (2,0,2)-mode may form a large mixing. 
    
    \begingroup
    \squeezetable
    \begin{table}[htbp]
        \caption{$LS$-coupling scheme in the three-body angular-momentum space for triply heavy baryons. Parameters $r_{\min}$, $r_{\max}$, $R_{\min}$, and $R_{\max}$ are in fm.\label{tab:Angular-momentum space}}
        \begin{ruledtabular}
        \begin{tabular}{cccccccccccc}
            $J^P$&$l$&$L$&$L_t$&$s_{ij}$&$S$&$n_{\max}$&$r_{\min}$&$r_{\max}$&$N_{\max}$&$R_{\min}$&$R_{\max}$\\
            \hline
            \multirow{8.5}{*}{$\dfrac{1}{2}^+$}
            &0&0&0&1&$\frac{1}{2}$&10&0.1&5.0&10&0.1&5.0\\
            &1&1&0&0&$\frac{1}{2}$&10&0.1&5.0&10&0.1&5.0\\
            \cline{2-12}
            &0&2&2&1&$\frac{3}{2}$&10&0.1&5.0&10&0.1&5.0\\
            &2&0&2&1&$\frac{3}{2}$&10&0.1&5.0&10&0.1&5.0\\
            \cline{2-12}
            &1&1&1&0&$\frac{1}{2}$&10&0.1&5.0&10&0.1&5.0\\
            &2&2&1&1&$\frac{1}{2}$&10&0.1&5.0&10&0.1&5.0\\
            \cline{2-12}
            &2&2&1&1&$\frac{3}{2}$&10&0.1&5.0&10&0.1&5.0\\
            \hline
            \multirow{14}{*}{$\dfrac{3}{2}^+$}
            &0&0&0&1&$\frac{3}{2}$&10&0.1&5.0&10&0.1&5.0\\
            &2&2&0&1&$\frac{3}{2}$&10&0.1&5.0&10&0.1&5.0\\
            \cline{2-12}
            &0&2&2&1&$\frac{3}{2}$&10&0.1&5.0&10&0.1&5.0\\
            &2&0&2&1&$\frac{3}{2}$&10&0.1&5.0&10&0.1&5.0\\
            \cline{2-12}
            &2&0&2&1&$\frac{1}{2}$&10&0.1&5.0&10&0.1&5.0\\
            &1&1&2&0&$\frac{1}{2}$&10&0.1&5.0&10&0.1&5.0\\
            &0&2&2&1&$\frac{1}{2}$&10&0.1&5.0&10&0.1&5.0\\
            \cline{2-12}
            &1&1&1&0&$\frac{1}{2}$&10&0.1&5.0&10&0.1&5.0\\
            &2&2&1&1&$\frac{1}{2}$&10&0.1&5.0&10&0.1&5.0\\
            \cline{2-12}
            &2&2&1&1&$\frac{3}{2}$&10&0.1&5.0&10&0.1&5.0\\
            \cline{2-12}
            &2&2&3&1&$\frac{3}{2}$&10&0.1&5.0&10&0.1&5.0\\
            \hline
            \multirow{12.5}{*}{$\dfrac{5}{2}^+$}
            &0&2&2&1&$\frac{3}{2}$&10&0.1&5.0&10&0.1&5.0\\
            &2&0&2&1&$\frac{3}{2}$&10&0.1&5.0&10&0.1&5.0\\
            \cline{2-12}
            &2&0&2&1&$\frac{1}{2}$&10&0.1&5.0&10&0.1&5.0\\
            &1&1&2&0&$\frac{1}{2}$&10&0.1&5.0&10&0.1&5.0\\
            &0&2&2&1&$\frac{1}{2}$&10&0.1&5.0&10&0.1&5.0\\
            \cline{2-12}
            &2&2&1&1&$\frac{3}{2}$&10&0.1&5.0&10&0.1&5.0\\
            \cline{2-12}
            &2&2&3&1&$\frac{3}{2}$&10&0.1&5.0&10&0.1&5.0\\
            \cline{2-12}
            &1&3&3&0&$\frac{1}{2}$&10&0.1&5.0&10&0.1&5.0\\
            &2&2&3&1&$\frac{1}{2}$&10&0.1&5.0&10&0.1&5.0\\
            &3&1&3&0&$\frac{1}{2}$&10&0.1&5.0&10&0.1&5.0\\
            \hline
            \multirow{7.3}{*}{$\dfrac{7}{2}^+$}
            &0&2&2&1&$\frac{3}{2}$&10&0.1&5.0&10&0.1&5.0\\
            &2&0&2&1&$\frac{3}{2}$&10&0.1&5.0&10&0.1&5.0\\
            \cline{2-12}
            &2&2&3&1&$\frac{3}{2}$&10&0.1&5.0&10&0.1&5.0\\
            \cline{2-12}
            &1&3&3&0&$\frac{1}{2}$&10&0.1&5.0&10&0.1&5.0\\
            &2&2&3&1&$\frac{1}{2}$&10&0.1&5.0&10&0.1&5.0\\
            &3&1&3&0&$\frac{1}{2}$&10&0.1&5.0&10&0.1&5.0\\
            \hline
            \multirow{4.6}{*}{$\dfrac{1}{2}^-$}
            &0&1&1&1&$\frac{1}{2}$&10&0.1&5.0&10&0.1&5.0\\
            &1&0&1&0&$\frac{1}{2}$&10&0.1&5.0&10&0.1&5.0\\
            \cline{2-12}
            &0&1&1&1&$\frac{3}{2}$&10&0.1&5.0&10&0.1&5.0\\
            \cline{2-12}
            &2&1&2&1&$\frac{3}{2}$&10&0.1&5.0&10&0.1&5.0\\
            \hline
            \multirow{10}{*}{$\dfrac{3}{2}^-$}
            &0&1&1&1&$\frac{1}{2}$&10&0.1&5.0&10&0.1&5.0\\
            &1&0&1&0&$\frac{1}{2}$&10&0.1&5.0&10&0.1&5.0\\
            \cline{2-12}
            &0&1&1&1&$\frac{3}{2}$&10&0.1&5.0&10&0.1&5.0\\
            \cline{2-12}
            &0&3&3&1&$\frac{3}{2}$&10&0.1&5.0&10&0.1&5.0\\
            &2&1&3&1&$\frac{3}{2}$&10&0.1&5.0&10&0.1&5.0\\
            \cline{2-12}
            &1&2&2&0&$\frac{1}{2}$&10&0.1&5.0&10&0.1&5.0\\
            &2&1&2&1&$\frac{1}{2}$&10&0.1&5.0&10&0.1&5.0\\
            \cline{2-12}
            &2&1&2&1&$\frac{3}{2}$&10&0.1&5.0&10&0.1&5.0\\
            \hline
            \multirow{14}{*}{$\dfrac{5}{2}^-$}
            &0&1&1&1&$\frac{3}{2}$&10&0.1&5.0&10&0.1&5.0\\
            \cline{2-12}
            &0&3&3&1&$\frac{1}{2}$&10&0.1&5.0&10&0.1&5.0\\
            &1&2&3&0&$\frac{1}{2}$&10&0.1&5.0&10&0.1&5.0\\
            &2&1&3&1&$\frac{1}{2}$&10&0.1&5.0&10&0.1&5.0\\
            &3&0&3&0&$\frac{1}{2}$&10&0.1&5.0&10&0.1&5.0\\
            \cline{2-12}
            &0&3&3&1&$\frac{3}{2}$&10&0.1&5.0&10&0.1&5.0\\
            &2&1&3&1&$\frac{3}{2}$&10&0.1&5.0&10&0.1&5.0\\
            \cline{2-12}
            &1&2&2&0&$\frac{1}{2}$&10&0.1&5.0&10&0.1&5.0\\
            &2&1&2&1&$\frac{1}{2}$&10&0.1&5.0&10&0.1&5.0\\
            \cline{2-12}
            &2&1&2&1&$\frac{3}{2}$&10&0.1&5.0&10&0.1&5.0\\
        \end{tabular}
        \end{ruledtabular}
    \end{table}
    \endgroup

    \subsection{Gaussian expansion method}

The system described by Hamiltonian $H$ satisfies the stationary Schr\"odinger equation
\begin{equation}
    H\Psi = E\Psi.
\end{equation}
We expand the total wave function in terms of a complete set of $L^2$-integrable basis functions $\Phi_\alpha$, where $\alpha$ denotes the set of quantum numbers labeling each basis state:
\begin{equation}
    \Psi = \sum_\alpha C_\alpha \Phi_\alpha.
\end{equation}
Applying the Rayleigh-Ritz variational principle yields a generalized matrix eigenvalue problem:
\begin{equation}
    HC = NCE,
\end{equation}
where the Hamiltonian and overlap matrix elements are respectively given by
\begin{align}
    H_{\alpha'\alpha} &= \langle \Phi_{\alpha'} | H | \Phi_\alpha \rangle, \label{eq:energy_matrix_element} \\
    N_{\alpha'\alpha} &= \langle \Phi_{\alpha'} | \Phi_\alpha \rangle.
\end{align}

The GEM employs carefully selected Gaussian basis functions that form an approximately complete set in finite coordinate space. This choice enables accurate description of both short-range correlations and long-range asymptotic behavior, as well as the highly oscillatory nature of wave functions in bound and scattering states \cite{Hiyama:2003cu}.



In Fig.~\ref{fig:Jacobian coordinates}, the spin-space basis for the three subfigures is defined as
\begin{equation}
\begin{split}
\phi_\alpha^{c=1}=&\left[[\phi_{nl}^{\mathrm{G}}(\boldsymbol{r}_1)\phi_{NL}^{\mathrm{G}}(\boldsymbol{R}_1)]_{L_t}[s_1[s_2s_3]_{s_{23}}]_S\right]_{JM},\\
\phi_\alpha^{c=2}=&\left[[\phi_{nl}^{\mathrm{G}}(\boldsymbol{r}_2)\phi_{NL}^{\mathrm{G}}(\boldsymbol{R}_2)]_{L_t}[s_2[s_3s_1]_{s_{31}}]_S\right]_{JM}, \\
\phi_\alpha^{c=3}=&\left[[\phi_{nl}^{\mathrm{G}}(\boldsymbol{r}_3)\phi_{NL}^{\mathrm{G}}(\boldsymbol{R}_3)]_{L_t}[s_3[s_1s_2]_{s_{12}}]_S\right]_{JM},
\end{split}
\end{equation}
where $\alpha = \{l, L, L_t, s_{ij}, S, n, N\}$ denotes the set of quantum numbers. For the $\Omega_{ccc}$ and $\Omega_{bbb}$ systems, accounting for the symmetry of three identical quarks, we construct the symmetric basis:
\begin{equation}\label{eq:Phi1}
\Phi_\alpha = \phi_\alpha^{c=1}+ \phi_\alpha^{c=2}+\phi_\alpha^{c=3}.\\
\end{equation}
To verify the symmetry of Eq.~\eqref{eq:Phi1}, we apply the quark exchange operator $\hat{P}_{12}$ to $\phi_\alpha^{c=1}$, which exchanges the states of quarks 1 and 2. This yields
\begin{equation}
\hat{P}_{12} \phi_\alpha^{c=1} = (-1)^{l+s_{23}-1}\phi_\alpha^{c=2}.
\end{equation}
Given that the flavor wave functions of $\Omega_{ccc}$ and $\Omega_{bbb}$ are symmetric while the color wave functions are antisymmetric, $l$ and $s_{23}$ must satisfy
\begin{equation}
(-1)^{l+s_{23}-1}=1, \quad \text{i.e.,} \quad l + s_{23} = \text{odd}.
\end{equation}
Consequently,
\begin{equation}
\hat{P}_{12} \phi_\alpha^{c=1} = \phi_\alpha^{c=2}.
\end{equation}
Similarly, we find
\begin{equation}
\hat{P}_{13} \phi_\alpha^{c=1} = \phi_\alpha^{c=3}.
\end{equation}
Crucially, the operation $\hat{P}_{ij}\phi_\alpha^c$ remains within the function space spanned by ${\phi_\alpha^{c=1}, \phi_\alpha^{c=2}, \phi_\alpha^{c=3}}$, leading to
\begin{equation}
\hat{P}_{ij}\Phi_\alpha=\Phi_\alpha,
\end{equation}
which demonstrates the symmetry of the spin-space wave function. Since the flavor wave function is symmetric and the color wave function is antisymmetric, the total wave function is antisymmetric. For the $\Omega_{bcc}$ and $\Omega_{bbc}$ systems, which contain two identical quarks at positions $m_2$ and $m_3$ in Fig.~\ref{fig:Jacobian coordinates}, we adopt the basis:
\begin{equation}\label{eq:Phi2}
\Phi_\alpha = \phi_\alpha^{c=1}.
\end{equation}%
The quantum numbers used for each $J^P$ state are listed in Table~\ref{tab:Angular-momentum space}. Here we simplify notation by abbreviating $l_c$ and $L_c$ to $l$ and $L$ (due to identical particle symmetry) and denote the total orbital angular momentum as $L_t$ to avoid confusion.

The spatial components of Eqs.~\eqref{eq:Phi1} and \eqref{eq:Phi2} employ Gaussian basis functions of the form:
\begin{equation}
    \begin{array}{ll}
        \phi_{nlm}^{\mathrm{G}}(\boldsymbol{r}) = \tilde{\phi}_{nl}^{\mathrm{G}}(r)Y_{lm}(\hat{\boldsymbol{r}}), & \tilde{\phi}_{nl}^{\mathrm{G}}(r) = N_{nl}r^l e^{-\nu_n r^2}, \\
        \phi_{NLM}^{\mathrm{G}}(\boldsymbol{R}) = \tilde{\phi}_{NL}^{\mathrm{G}}(R)Y_{LM}(\hat{\boldsymbol{R}}), & \tilde{\phi}_{NL}^{\mathrm{G}}(R) = N_{NL}R^L e^{-\lambda_N R^2},
    \end{array}
\end{equation}
with normalization constants:
\begin{equation}
    N_{nl} = \left(\frac{2^{l+2}(2\nu_n)^{l+\frac{3}{2}}}{\sqrt{\pi}(2l+1)!!}\right)^{\frac{1}{2}}, \quad
    N_{NL} = \left(\frac{2^{L+2}(2\lambda_N)^{L+\frac{3}{2}}}{\sqrt{\pi}(2L+1)!!}\right)^{\frac{1}{2}},
\end{equation}
and Gaussian range parameters:
\begin{equation}
    \begin{array}{lll}
        \nu_n = 1/r_n^2, & r_n = r_1 a^{n-1} & (n = 1 \sim n_{\max}), \\
        \lambda_N = 1/R_N^2, & R_N = R_1 A^{N-1} & (N = 1 \sim N_{\max}).
    \end{array}
\end{equation}
Empirical evidence suggests that geometric progression provides optimal Gaussian size parameters.

The basis functions $\Phi_\alpha$ are constrained by:
\begin{itemize}
    \item Total angular momentum and parity conservation ($J^P$),
    \item Symmetrization requirements for identical particles,
    \item Hamiltonian commutation relations.
\end{itemize}

For given $J^P$, $\Phi_\alpha$ must satisfy:
\begin{itemize}
    \item Angular momentum coupling: $|L_t - S| \leq J \leq L_t + S$,
    \item Parity: $P = (-1)^{l+L}$,
    \item Symmetry condition: $l + s_{ij} = \text{odd}$ (from symmetrization requirements).
\end{itemize}

The tensor and spin-orbit terms leading to the mixing of different $L_t$ and $S$ can be expressed as irreducible tensors:
\begin{align}
    S(\boldsymbol{r}_{ij},\boldsymbol{S}_{i},\boldsymbol{S}_j) &= \sqrt{24\pi} \left(Y_2(\hat{\boldsymbol{r}}_{ij}) \otimes (S_i^{(1)} \otimes S_j^{(1)})^{(2)}\right)^{(0)}_0, \\
    \boldsymbol{L\cdot S} &= -\sqrt{3} (L^{(1)} \otimes S^{(1)})^{(0)}_0,
\end{align}
where $\boldsymbol{L}$ and $\boldsymbol{S}$ correspond to any $\boldsymbol{r}\times\boldsymbol{p}$ and $\boldsymbol{S}_k$ in $V_{ij}^{\text{so(cm)}}$ and $V_{ij}^{\text{so(tp)}}$.

The energy matrix elements (Eq.~\eqref{eq:energy_matrix_element}) decompose into spatial and spin components via:
\begin{equation}
    \begin{aligned}
        &\langle j_1'j_2'j'm'|(T_1^{(k_1)} \otimes T_2^{(k_2)})^{(k)}_q|j_1j_2jm\rangle = \\
        &(-1)^{j'-m'} \begin{pmatrix}
            j' & k & j \\
            -m' & q & m
        \end{pmatrix}
        \sqrt{(2j'+1)(2k+1)(2j+1)} \\
        &\begin{Bmatrix}
            j_1' & j_1 & k_1 \\
            j_2' & j_2 & k_2 \\
            j' & j & k
        \end{Bmatrix}
        \langle j_1' \| T_1^{(k_1)} \| j_1 \rangle \langle j_2' \| T_2^{(k_2)} \| j_2 \rangle,
    \end{aligned}
\end{equation}
where $(j_1'k_1j_1)$ and $(j_2'k_2j_2)$ must satisfy triangular conditions.

For different terms, non-vanishing matrix elements require:
\begin{itemize}
    \item Tensor/spin-orbit terms: $|L_t'-L_t| \leq 2/1 \leq L_t'+L_t$ and $|S'-S| \leq 2/1 \leq S'+S$,
    \item Commuting terms with $L^2$ and $S^2$: $L_t' = L_t$ and $S' = S$.
\end{itemize}

In practice, we consider only $L_t \leq 3$ for $J^P = \frac{1}{2}^\pm, \frac{3}{2}^\pm, \frac{5}{2}^\pm, \frac{7}{2}^+$, neglecting higher orbital excitations. Table~\ref{tab:Angular-momentum space} summarizes the allowed quantum numbers and Gaussian ranges for each $J^P$ state of triply heavy baryons.

    \subsection{Radiative decay formalism}
    
    Using the obtained wave functions of triply heavy baryons, we can calculate their radiative decay properties. The theoretical framework for calculating radiative decays within the nonrelativistic constituent quark model has been well established in Refs.~\cite{Deng:2015bva,Deng:2016stx,Deng:2016ktl}, and this formalism has been successfully applied to various systems \cite{Peng:2024pyl}. Below we briefly summarize the key steps of the calculation.

The quark-photon electromagnetic interaction at tree level is described by
\begin{equation}
    H_e = -\sum_j e_j \bar{\psi}_j \gamma_\mu^j A^\mu(\boldsymbol{k},\boldsymbol{r}) \psi_j,
\end{equation}
where $\psi_j$ represents the field of the $j$-th quark in the hadron, $e_j$ and $\gamma^j_\mu$ denote the electric charge and Dirac matrices of the constituent quark $\psi_j$, and $A^\mu(\boldsymbol{k},\boldsymbol{r})$ is the photon field.

In the nonrelativistic limit, this interaction simplifies to
\begin{equation}\label{eq:he}
    h_e \approx \sum_{j} \left[ e_{j} \boldsymbol{r}_{j} \cdot \boldsymbol{\epsilon} 
    - \frac{e_{j}}{2m_{j}} \boldsymbol{\sigma}_{j} \cdot (\boldsymbol{\epsilon} \times \hat{\boldsymbol{k}}) \right] 
    \mathrm{e}^{-\mathrm{i}\boldsymbol{k} \cdot \boldsymbol{r}_j},
\end{equation}
where $m_j$, $e_j$, $\boldsymbol{\sigma}_j$, and $\boldsymbol{r}_j$ are the mass, charge, Pauli matrices, and position of the $j$-th quark, respectively. The $\boldsymbol{k}$ and $\boldsymbol{\epsilon}$ in Eq.~\eqref{eq:he} represent the momentum and polarization vector of the photon, respectively.

The corresponding helicity amplitude for the electromagnetic transition is given by
\begin{equation}\label{eq:AmpA}
    \mathcal{A} = -\mathrm{i} \sqrt{\frac{\omega_\gamma}{2}} \langle f | h_e | i \rangle,
\end{equation}
where $\langle f|$ and $|i\rangle$ denote the wave functions of the final and initial baryon states, respectively. The $\omega_\gamma = (M_i^2 - M_f^2)/(2M_i)$ in Eq.~\eqref{eq:AmpA} is the photon energy, where $M_i$ and $M_f$ are the masses of the initial and final baryons, respectively. We choose the coordinate system such that the photon momentum aligns with the $z$-axis ($\boldsymbol{k} = (0,0,\omega_\gamma)$) and take the polarization vector as $\boldsymbol{\epsilon} = (1,\mathrm{i},0)/\sqrt{2}$.

The radiative decay width is then obtained from the helicity amplitude through
\begin{equation}
    \Gamma = \frac{|\boldsymbol{k}|^2}{\pi} \frac{2}{2J_i+1} \frac{M_f}{M_i} 
    \sum_{J_{fz},J_{iz}} |\mathcal{A}_{J_{fz},J_{iz}}|^2,
\end{equation}
where $J_i$ is the total angular momentum of the initial baryon, and $J_{iz}$ and $J_{fz}$ are the $z$-components of the angular momenta for the initial and final states, respectively.
    
    \section{Numerical RESULTS}\label{sec3}

\subsection{Mass spectra}
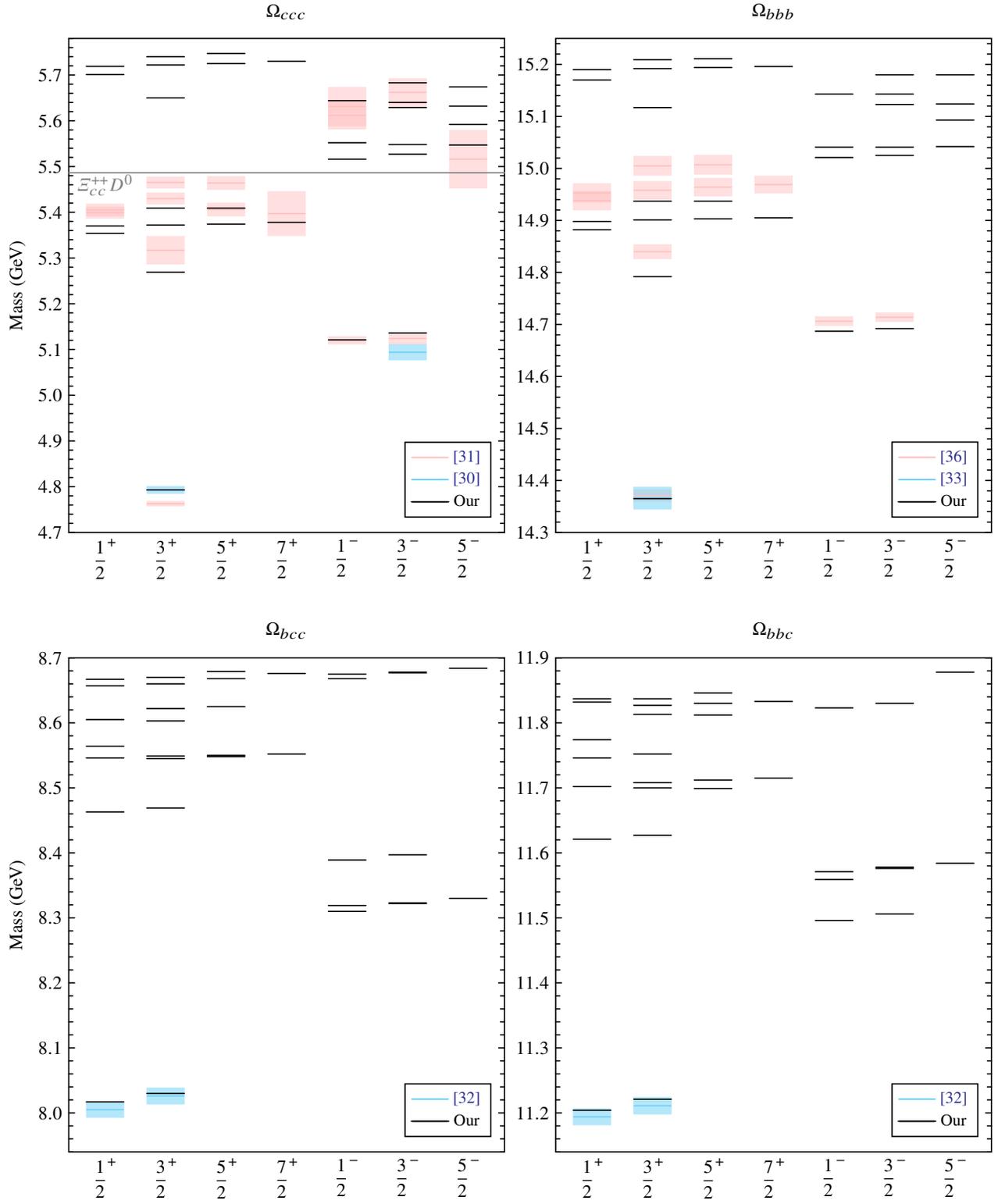
\begin{figure*}[htbp]
        \pgfplotsset{
            /pgfplots/error bars/draw error bar/.code 2 args={
                \pgfkeysgetvalue{/pgfplots/error bars/error mark}
                {\pgfplotserrorbarsmark}
                \pgfkeysgetvalue{/pgfplots/error bars/error mark options}
                {\pgfplotserrorbarsmarkopts}
                \draw[line width=0.65cm, opacity=0.5] #1-- #2 node [pos=1,sloped,allow upside down] {
                    \expandafter\tikz\expandafter[\pgfplotserrorbarsmarkopts]{
                        \expandafter\pgfuseplotmark\expandafter{}
                        \pgfusepath{stroke}}
                };
            },
        }
        \begin{tikzpicture}
            \begin{axis}[
                title=$\Omega_{ccc}$,
                width=9cm, height=10cm,
                axis line style={semithick},
                ylabel=Mass (GeV),
                ymin=4.7, ymax=5.78,
                ytick distance=0.1, minor y tick num=4,
                yticklabel style={/pgf/number format/.cd, fixed, precision=1, fixed zerofill},
                every y tick/.append style={semithick, black, line cap=round},
                major tick length=0.1cm,
                minor tick length=0.07cm,
                symbolic x coords={0,$\dfrac{1}{2}^+$,$\dfrac{3}{2}^+$,$\dfrac{5}{2}^+$,$\dfrac{7}{2}^+$,$\dfrac{1}{2}^-$,$\dfrac{3}{2}^-$,$\dfrac{5}{2}^-$,8},
                xtick style={draw=none},
                legend entries={\cite{Padmanath:2013zfa},\cite{Dhindsa:2024erk},Our},
                legend pos=south east,
                legend style={semithick, font=\footnotesize},
                ]
                \addplot[pink, draw=none,
                mark=-, mark size=0.32cm, semithick,
                error bars/.cd, y dir=both, y explicit,
                ]coordinates
                {
                    ($\dfrac{1}{2}^+$,5.399)+-(0,0.013)
                    ($\dfrac{1}{2}^+$,5.405)+-(0,0.014)
                    ($\dfrac{3}{2}^+$,4.763)+-(0,0.006)
                    ($\dfrac{3}{2}^+$,5.317)+-(0,0.031)
                    ($\dfrac{3}{2}^+$,5.430)+-(0,0.013)
                    ($\dfrac{3}{2}^+$,5.465)+-(0,0.013)
                    ($\dfrac{5}{2}^+$,5.406)+-(0,0.015)
                    ($\dfrac{5}{2}^+$,5.464)+-(0,0.015)
                    ($\dfrac{7}{2}^+$,5.397)+-(0,0.049)
                    ($\dfrac{1}{2}^-$,5.120)+-(0,0.009)
                    ($\dfrac{1}{2}^-$,5.612)+-(0,0.031)
                    ($\dfrac{1}{2}^-$,5.631)+-(0,0.043)
                    ($\dfrac{3}{2}^-$,5.124)+-(0,0.013)
                    ($\dfrac{3}{2}^-$,5.662)+-(0,0.031)
                    ($\dfrac{5}{2}^-$,5.516)+-(0,0.064)
                };
                \addplot[cyan!50, draw=none,
                mark=-, mark size=0.32cm, semithick,
                error bars/.cd, y dir=both, y explicit,
                ]coordinates
                {
                    ($\dfrac{3}{2}^+$,4.793)+-(0,0.0086)
                    ($\dfrac{3}{2}^-$,5.094)+-(0,0.0177)
                };
                \addplot[mark=-, mark size=0.32cm, semithick, line cap=round,
                draw=none, point meta=explicit symbolic,
                coordinate style/.condition={y==5.323||y==5.613||y==5.473}{below=-2.5pt},
                ]coordinates
                {
                    ($\dfrac{1}{2}^+$,5.354)[5354]
                    ($\dfrac{1}{2}^+$,5.370)[5370]
                    ($\dfrac{1}{2}^+$,5.701)[5701]
                    ($\dfrac{1}{2}^+$,5.719)[5719]
                    ($\dfrac{1}{2}^+$,5.784)[5784]
                    ($\dfrac{1}{2}^+$,5.851)[5851]
                    ($\dfrac{1}{2}^+$,5.885)[5885]
                    ($\dfrac{1}{2}^+$,6.011)[6011]
                    ($\dfrac{3}{2}^+$,4.793)[4793]
                    ($\dfrac{3}{2}^+$,5.269)[5269]
                    ($\dfrac{3}{2}^+$,5.372)[5372]
                    ($\dfrac{3}{2}^+$,5.409)[5409]
                    ($\dfrac{3}{2}^+$,5.650)[5650]
                    ($\dfrac{3}{2}^+$,5.722)[5722]
                    ($\dfrac{3}{2}^+$,5.740)[5740]
                    ($\dfrac{3}{2}^+$,5.796)[5796]
                    ($\dfrac{5}{2}^+$,5.374)[5374]
                    ($\dfrac{5}{2}^+$,5.409)[5409]
                    ($\dfrac{5}{2}^+$,5.725)[5725]
                    ($\dfrac{5}{2}^+$,5.747)[5747]
                    ($\dfrac{5}{2}^+$,5.801)[5801]
                    ($\dfrac{5}{2}^+$,5.852)[5852]
                    ($\dfrac{5}{2}^+$,5.883)[5883]
                    ($\dfrac{5}{2}^+$,5.887)[5887]
                    ($\dfrac{7}{2}^+$,5.378)[5378]
                    ($\dfrac{7}{2}^+$,5.730)[5730]
                    ($\dfrac{7}{2}^+$,5.879)[5879]
                    ($\dfrac{7}{2}^+$,5.891)[5891]
                    ($\dfrac{7}{2}^+$,6.041)[6041]
                    ($\dfrac{7}{2}^+$,6.131)[6131]
                    ($\dfrac{7}{2}^+$,6.186)[6186]
                    ($\dfrac{7}{2}^+$,6.200)[6200]
                    ($\dfrac{1}{2}^-$,5.121)[5121]
                    ($\dfrac{1}{2}^-$,5.516)[5516]
                    ($\dfrac{1}{2}^-$,5.552)[5552]
                    ($\dfrac{1}{2}^-$,5.644)[5644]
                    ($\dfrac{1}{2}^-$,5.854)[5854]
                    ($\dfrac{1}{2}^-$,5.872)[5872]
                    ($\dfrac{1}{2}^-$,5.958)[5958]
                    ($\dfrac{1}{2}^-$,5.983)[5983]
                    ($\dfrac{3}{2}^-$,5.136)[5136]
                    ($\dfrac{3}{2}^-$,5.527)[5527]
                    ($\dfrac{3}{2}^-$,5.548)[5548]
                    ($\dfrac{3}{2}^-$,5.629)[5629]
                    ($\dfrac{3}{2}^-$,5.640)[5640]
                    ($\dfrac{3}{2}^-$,5.683)[5683]
                    ($\dfrac{3}{2}^-$,5.862)[5862]
                    ($\dfrac{3}{2}^-$,5.868)[5868]
                    ($\dfrac{5}{2}^-$,5.547)[5547]
                    ($\dfrac{5}{2}^-$,5.592)[5592]
                    ($\dfrac{5}{2}^-$,5.632)[5632]
                    ($\dfrac{5}{2}^-$,5.674)[5674]
                    ($\dfrac{5}{2}^-$,5.868)[5868]
                    ($\dfrac{5}{2}^-$,5.910)[5910]
                    ($\dfrac{5}{2}^-$,5.930)[5930]
                    ($\dfrac{5}{2}^-$,5.997)[5997]
                    
                };
                \addplot[gray, semithick, line legend,
                sharp plot, update limits=false,
                ]coordinates {(0,5.48644)(8,5.48644)}
                node at (0,76) {$\varXi_{cc}^{++}D^0$};
            \end{axis}
        \end{tikzpicture}
        \begin{tikzpicture}
            \begin{axis}[
                title=$\Omega_{bbb}$,
                width=9cm, height=10cm,
                axis line style={semithick},
                ymin=14.3, ymax=15.25,
                ytick distance=0.1, minor y tick num=4,
                yticklabel style={/pgf/number format/.cd, fixed, precision=1, fixed zerofill},
                every y tick/.append style={semithick, black, line cap=round},
                major tick length=0.1cm,
                minor tick length=0.07cm,
                symbolic x coords={$\dfrac{1}{2}^+$,$\dfrac{3}{2}^+$,$\dfrac{5}{2}^+$,$\dfrac{7}{2}^+$,$\dfrac{1}{2}^-$,$\dfrac{3}{2}^-$,$\dfrac{5}{2}^-$},
                xtick style={draw=none},
                legend entries={\cite{Meinel:2012qz},\cite{Brown:2014ena},Our},
                legend pos=south east,
                legend style={semithick, font=\footnotesize},
                ]
                \addplot[pink, draw=none,
                mark=-, mark size=0.32cm, semithick,
                error bars/.cd, y dir=both, y explicit,
                ]coordinates
                {
                    ($\dfrac{1}{2}^+$,14.938)+-(0,0.0184)
                    ($\dfrac{1}{2}^+$,14.953)+-(0,0.0184)
                    ($\dfrac{3}{2}^+$,14.371)+-(0,0.0117)
                    ($\dfrac{3}{2}^+$,14.840)+-(0,0.0142)
                    ($\dfrac{3}{2}^+$,14.958)+-(0,0.0177)
                    ($\dfrac{3}{2}^+$,15.005)+-(0,0.0191)
                    ($\dfrac{5}{2}^+$,14.964)+-(0,0.0177)
                    ($\dfrac{5}{2}^+$,15.007)+-(0,0.0191)
                    ($\dfrac{7}{2}^+$,14.969)+-(0,0.0170)
                    ($\dfrac{1}{2}^-$,14.7063)+-(0,0.0094)
                    ($\dfrac{3}{2}^-$,14.7140)+-(0,0.0091)
                };
                \addplot[cyan!50, draw=none,
                mark=-, mark size=0.32cm, semithick,
                error bars/.cd, y dir=both, y explicit,
                ]coordinates
                {
                    ($\dfrac{3}{2}^+$,14.366)+-(0,0.0219)
                };
                \addplot[mark=-, mark size=0.32cm, semithick, line cap=round,
                draw=none, point meta=explicit symbolic,
                coordinate style/.condition={y==5.323||y==5.613||y==5.473}{below=-2.5pt},
                ]coordinates
                {
                    ($\dfrac{1}{2}^+$,14.882)[14882]
                    ($\dfrac{1}{2}^+$,14.898)[14898]
                    ($\dfrac{1}{2}^+$,15.170)[15170]
                    ($\dfrac{1}{2}^+$,15.190)[15190]
                    ($\dfrac{1}{2}^+$,15.257)[15257]
                    ($\dfrac{1}{2}^+$,15.324)[15324]
                    ($\dfrac{1}{2}^+$,15.349)[15349]
                    ($\dfrac{1}{2}^+$,15.424)[15424]
                    ($\dfrac{3}{2}^+$,14.365)[14365]
                    ($\dfrac{3}{2}^+$,14.792)[14792]
                    ($\dfrac{3}{2}^+$,14.901)[14901]
                    ($\dfrac{3}{2}^+$,14.937)[14937]
                    ($\dfrac{3}{2}^+$,15.117)[15117]
                    ($\dfrac{3}{2}^+$,15.192)[15192]
                    ($\dfrac{3}{2}^+$,15.209)[15209]
                    ($\dfrac{3}{2}^+$,15.266)[15266]
                    ($\dfrac{5}{2}^+$,14.903)[14903]
                    ($\dfrac{5}{2}^+$,14.937)[14937]
                    ($\dfrac{5}{2}^+$,15.194)[15194]
                    ($\dfrac{5}{2}^+$,15.211)[15211]
                    ($\dfrac{5}{2}^+$,15.267)[15267]
                    ($\dfrac{5}{2}^+$,15.327)[15327]
                    ($\dfrac{5}{2}^+$,15.351)[15351]
                    ($\dfrac{5}{2}^+$,15.354)[15354]
                    ($\dfrac{7}{2}^+$,14.905)[14905]
                    ($\dfrac{7}{2}^+$,15.196)[15196]
                    ($\dfrac{7}{2}^+$,15.352)[15352]
                    ($\dfrac{7}{2}^+$,15.356)[15356]
                    ($\dfrac{7}{2}^+$,15.448)[15448]
                    ($\dfrac{7}{2}^+$,15.538)[15538]
                    ($\dfrac{7}{2}^+$,15.600)[15600]
                    ($\dfrac{7}{2}^+$,15.608)[15608]
                    ($\dfrac{1}{2}^-$,14.687)[14687]
                    ($\dfrac{1}{2}^-$,15.021)[15021]
                    ($\dfrac{1}{2}^-$,15.041)[15041]
                    ($\dfrac{1}{2}^-$,15.143)[15143]
                    ($\dfrac{1}{2}^-$,15.299)[15299]
                    ($\dfrac{1}{2}^-$,15.304)[15304]
                    ($\dfrac{1}{2}^-$,15.398)[15398]
                    ($\dfrac{1}{2}^-$,15.422)[15422]
                    ($\dfrac{3}{2}^-$,14.692)[14692]
                    ($\dfrac{3}{2}^-$,15.025)[15025]
                    ($\dfrac{3}{2}^-$,15.041)[15041]
                    ($\dfrac{3}{2}^-$,15.123)[15123]
                    ($\dfrac{3}{2}^-$,15.143)[15143]
                    ($\dfrac{3}{2}^-$,15.180)[15180]
                    ($\dfrac{3}{2}^-$,15.302)[15302]
                    ($\dfrac{3}{2}^-$,15.304)[15304]
                    ($\dfrac{5}{2}^-$,15.042)[15042]
                    ($\dfrac{5}{2}^-$,15.093)[15093]
                    ($\dfrac{5}{2}^-$,15.124)[15124]
                    ($\dfrac{5}{2}^-$,15.180)[15180]
                    ($\dfrac{5}{2}^-$,15.305)[15305]
                    ($\dfrac{5}{2}^-$,15.352)[15352]
                    ($\dfrac{5}{2}^-$,15.364)[15364]
                    ($\dfrac{5}{2}^-$,15.436)[15436]
                    
                };
            \end{axis}
        \end{tikzpicture}\\[0.5cm]
        \begin{tikzpicture}
            \begin{axis}[
                title=$\Omega_{bcc}$,
                width=9cm, height=10cm,
                axis line style={semithick},
                ylabel=Mass (GeV),
                ymin=7.94, ymax=8.7,
                ytick distance=0.1, minor y tick num=4,
                yticklabel style={/pgf/number format/.cd, fixed, precision=1, fixed zerofill},
                every y tick/.append style={semithick, black, line cap=round},
                major tick length=0.1cm,
                minor tick length=0.07cm,
                symbolic x coords={$\dfrac{1}{2}^+$,$\dfrac{3}{2}^+$,$\dfrac{5}{2}^+$,$\dfrac{7}{2}^+$,$\dfrac{1}{2}^-$,$\dfrac{3}{2}^-$,$\dfrac{5}{2}^-$},
                xtick style={draw=none},
                legend entries={\cite{Mathur:2018epb},Our},
                legend pos=south east,
                legend style={semithick, font=\footnotesize},
                ]
                \addplot[cyan!50, draw=none,
                mark layer=axis foreground,
                mark=-, mark size=0.32cm, semithick,
                error bars/.cd, y dir=both, y explicit,
                ]coordinates
                {
                    ($\dfrac{1}{2}^+$,8.005)+-(0,0.0125)
                    ($\dfrac{3}{2}^+$,8.026)+-(0,0.013)
                };
                \addplot[mark=-, mark size=0.32cm, semithick, line cap=round,
                draw=none, point meta=explicit symbolic,
                coordinate style/.condition={y==5.323||y==5.613||y==5.473}{below=-2.5pt},
                ]coordinates
                {
                    ($\dfrac{1}{2}^+$,8.017)[8017]
                    ($\dfrac{1}{2}^+$,8.463)[8463]
                    ($\dfrac{1}{2}^+$,8.546)[8546]
                    ($\dfrac{1}{2}^+$,8.564)[8564]
                    ($\dfrac{1}{2}^+$,8.605)[8605]
                    ($\dfrac{1}{2}^+$,8.657)[8657]
                    ($\dfrac{1}{2}^+$,8.667)[8667]
                    ($\dfrac{1}{2}^+$,8.825)[8825]
                    ($\dfrac{3}{2}^+$,8.030)[8030]
                    ($\dfrac{3}{2}^+$,8.469)[8469]
                    ($\dfrac{3}{2}^+$,8.545)[8545]
                    ($\dfrac{3}{2}^+$,8.549)[8549]
                    ($\dfrac{3}{2}^+$,8.603)[8603]
                    ($\dfrac{3}{2}^+$,8.622)[8622]
                    ($\dfrac{3}{2}^+$,8.660)[8660]
                    ($\dfrac{3}{2}^+$,8.670)[8670]
                    ($\dfrac{5}{2}^+$,8.548)[8548]
                    ($\dfrac{5}{2}^+$,8.550)[8550]
                    ($\dfrac{5}{2}^+$,8.625)[8625]
                    ($\dfrac{5}{2}^+$,8.668)[8668]
                    ($\dfrac{5}{2}^+$,8.679)[8679]
                    ($\dfrac{5}{2}^+$,8.862)[8862]
                    ($\dfrac{5}{2}^+$,8.863)[8863]
                    ($\dfrac{5}{2}^+$,8.957)[8957]
                    ($\dfrac{7}{2}^+$,8.552)[8552]
                    ($\dfrac{7}{2}^+$,8.676)[8676]
                    ($\dfrac{7}{2}^+$,8.866)[8866]
                    ($\dfrac{7}{2}^+$,8.985)[8985]
                    ($\dfrac{7}{2}^+$,9.029)[9029]
                    ($\dfrac{7}{2}^+$,9.033)[9033]
                    ($\dfrac{7}{2}^+$,9.080)[9080]
                    ($\dfrac{7}{2}^+$,9.087)[9087]
                    ($\dfrac{1}{2}^-$,8.310)[8310]
                    ($\dfrac{1}{2}^-$,8.319)[8319]
                    ($\dfrac{1}{2}^-$,8.389)[8389]
                    ($\dfrac{1}{2}^-$,8.668)[8668]
                    ($\dfrac{1}{2}^-$,8.675)[8675]
                    ($\dfrac{1}{2}^-$,8.771)[8771]
                    ($\dfrac{1}{2}^-$,8.809)[8809]
                    ($\dfrac{1}{2}^-$,8.813)[8813]
                    ($\dfrac{3}{2}^-$,8.322)[8322]
                    ($\dfrac{3}{2}^-$,8.323)[8323]
                    ($\dfrac{3}{2}^-$,8.397)[8397]
                    ($\dfrac{3}{2}^-$,8.677)[8677]
                    ($\dfrac{3}{2}^-$,8.678)[8678]
                    ($\dfrac{3}{2}^-$,8.746)[8746]
                    ($\dfrac{3}{2}^-$,8.777)[8777]
                    ($\dfrac{3}{2}^-$,8.808)[8808]
                    ($\dfrac{5}{2}^-$,8.330)[8330]
                    ($\dfrac{5}{2}^-$,8.684)[8684]
                    ($\dfrac{5}{2}^-$,8.743)[8743]
                    ($\dfrac{5}{2}^-$,8.747)[8747]
                    ($\dfrac{5}{2}^-$,8.811)[8811]
                    ($\dfrac{5}{2}^-$,8.818)[8818]
                    ($\dfrac{5}{2}^-$,8.856)[8856]
                    ($\dfrac{5}{2}^-$,8.863)[8863]
                    
                };
                \addplot[mark=-, gray!50, mark size=0.32cm, semithick, line cap=round,
                draw=none, point meta=explicit symbolic,
                ]coordinates
                {
                    
                };
            \end{axis}
        \end{tikzpicture}
        \begin{tikzpicture}
            \begin{axis}[
                title=$\Omega_{bbc}$,
                width=9cm, height=10cm,
                axis line style={semithick},
                ymin=11.14, ymax=11.9,
                ytick distance=0.1, minor y tick num=4,
                yticklabel style={/pgf/number format/.cd, fixed, precision=1, fixed zerofill},
                every y tick/.append style={semithick, black, line cap=round},
                major tick length=0.1cm,
                minor tick length=0.07cm,
                symbolic x coords={$\dfrac{1}{2}^+$,$\dfrac{3}{2}^+$,$\dfrac{5}{2}^+$,$\dfrac{7}{2}^+$,$\dfrac{1}{2}^-$,$\dfrac{3}{2}^-$,$\dfrac{5}{2}^-$},
                xtick style={draw=none},
                legend entries={\cite{Mathur:2018epb},Our},
                legend pos=south east,
                legend style={semithick, font=\footnotesize},
                ]
                \addplot[cyan!50, draw=none,
                mark=-, mark size=0.32cm, semithick,
                error bars/.cd, y dir=both, y explicit,
                ]coordinates
                {
                    ($\dfrac{1}{2}^+$,11.194)+-(0,0.013)
                    ($\dfrac{3}{2}^+$,11.211)+-(0,0.0134)
                };
                \addplot[mark=-, mark size=0.32cm, semithick, line cap=round,
                draw=none, point meta=explicit symbolic,
                coordinate style/.condition={y==5.323||y==5.613||y==5.473}{below=-2.5pt},
                coordinate style/.condition={y==11.203}{red},
                ]coordinates
                {
                    ($\dfrac{1}{2}^+$,11.204)[11204]
                    ($\dfrac{1}{2}^+$,11.621)[11621]
                    ($\dfrac{1}{2}^+$,11.702)[11702]
                    ($\dfrac{1}{2}^+$,11.746)[11746]
                    ($\dfrac{1}{2}^+$,11.774)[11774]
                    ($\dfrac{1}{2}^+$,11.832)[11832]
                    ($\dfrac{1}{2}^+$,11.837)[11837]
                    ($\dfrac{1}{2}^+$,11.948)[11948]
                    ($\dfrac{3}{2}^+$,11.221)[11221]
                    ($\dfrac{3}{2}^+$,11.627)[11627]
                    ($\dfrac{3}{2}^+$,11.700)[11700]
                    ($\dfrac{3}{2}^+$,11.708)[11708]
                    ($\dfrac{3}{2}^+$,11.752)[11752]
                    ($\dfrac{3}{2}^+$,11.813)[11813]
                    ($\dfrac{3}{2}^+$,11.827)[11827]
                    ($\dfrac{3}{2}^+$,11.837)[11837]
                    ($\dfrac{5}{2}^+$,11.699)[11699]
                    ($\dfrac{5}{2}^+$,11.712)[11712]
                    ($\dfrac{5}{2}^+$,11.812)[11812]
                    ($\dfrac{5}{2}^+$,11.830)[11830]
                    ($\dfrac{5}{2}^+$,11.846)[11846]
                    ($\dfrac{5}{2}^+$,11.986)[11986]
                    ($\dfrac{5}{2}^+$,11.996)[11996]
                    ($\dfrac{5}{2}^+$,12.091)[12091]
                    ($\dfrac{7}{2}^+$,11.715)[11715]
                    ($\dfrac{7}{2}^+$,11.833)[11833]
                    ($\dfrac{7}{2}^+$,11.998)[11998]
                    ($\dfrac{7}{2}^+$,12.160)[12160]
                    ($\dfrac{7}{2}^+$,12.174)[12174]
                    ($\dfrac{7}{2}^+$,12.189)[12189]
                    ($\dfrac{7}{2}^+$,12.198)[12198]
                    ($\dfrac{7}{2}^+$,12.236)[12236]
                    ($\dfrac{1}{2}^-$,11.496)[11496]
                    ($\dfrac{1}{2}^-$,11.559)[11559]
                    ($\dfrac{1}{2}^-$,11.571)[11571]
                    ($\dfrac{1}{2}^-$,11.823)[11823]
                    ($\dfrac{1}{2}^-$,11.915)[11915]
                    ($\dfrac{1}{2}^-$,11.919)[11919]
                    ($\dfrac{1}{2}^-$,11.971)[11971]
                    ($\dfrac{1}{2}^-$,11.977)[11977]
                    ($\dfrac{3}{2}^-$,11.506)[11506]
                    ($\dfrac{3}{2}^-$,11.576)[11576]
                    ($\dfrac{3}{2}^-$,11.578)[11578]
                    ($\dfrac{3}{2}^-$,11.830)[11830]
                    ($\dfrac{3}{2}^-$,11.922)[11922]
                    ($\dfrac{3}{2}^-$,11.928)[11928]
                    ($\dfrac{3}{2}^-$,11.980)[11980]
                    ($\dfrac{3}{2}^-$,11.981)[11981]
                    ($\dfrac{5}{2}^-$,11.584)[11584]
                    ($\dfrac{5}{2}^-$,11.878)[11878]
                    ($\dfrac{5}{2}^-$,11.927)[11927]
                    ($\dfrac{5}{2}^-$,11.975)[11975]
                    ($\dfrac{5}{2}^-$,11.982)[11982]
                    ($\dfrac{5}{2}^-$,11.987)[11987]
                    ($\dfrac{5}{2}^-$,12.014)[12014]
                    ($\dfrac{5}{2}^-$,12.018)[12018]
                    
                };
                \addplot[mark=-, gray!50, mark size=0.32cm, semithick, line cap=round,
                draw=none, point meta=explicit symbolic,
                ]coordinates
                {
                    
                };
            \end{axis}
        \end{tikzpicture}
        \caption{The mass spectrum of triply heavy baryons. Here, blue and pink are the lattice results. \label{fig:Mass spectrum}}
    \end{figure*}

Using the GEM to numerically solve the stationary Schr\"odinger equation for triply heavy baryons, we obtain the mass spectra shown in Fig.~\ref{fig:Mass spectrum}, where we compare our results with lattice QCD calculations. The numerical values of the masses and angular momentum excitation proportions are presented in Tables~\ref{tab:ccc bbb Mass spectrum} and \ref{tab:bcc bbc Mass spectrum}, with lattice results provided in Table~\ref{tab:Lattice results} for reference.

In the absence of experimental data, we compare our results with lattice QCD calculations. For $\Omega_{ccc}$ baryons, most lattice studies have focused on the low-lying states with $J^P=\frac{3}{2}^+$ and $\frac{3}{2}^-$, with only Ref.~\cite{Padmanath:2013zfa} providing spectra up to $J^P=\frac{7}{2}^+$. For the $\frac{3}{2}^\pm$ states, we use the most recent lattice results from Ref.~\cite{Dhindsa:2024erk}, which claims to fully account for systematic uncertainties, while other $J^P$ states are compared with Ref.~\cite{Padmanath:2013zfa}. 

For $\Omega_{bcc}$ and $\Omega_{bbc}$ baryons, only three lattice studies \cite{Mathur:2018epb,Brown:2014ena,Burch:2015pka} have calculated masses for the low-lying $J^P=\frac{1}{2}^\pm$ and $\frac{3}{2}^\pm$ states (Table~\ref{tab:Lattice results}). In Fig.~\ref{fig:Mass spectrum}, we plot only the results from Ref.~\cite{Mathur:2018epb} due to their consistency with other calculations. For $\Omega_{bbb}$ baryons, the most recent published results \cite{Brown:2014ena} are a decade old, with highly excited states reported only in Refs.~\cite{Meinel:2010pw,Meinel:2012qz}. We omit the results from Ref.~\cite{Burch:2015pka} from the figure due to their larger uncertainties, though they are included in Table~\ref{tab:Lattice results} for completeness.

    \begingroup
    \squeezetable
    \begin{table*}[htbp]
        \caption{The mass spectra of the $\Omega_{ccc}$ and $\Omega_{bbb}$ baryons and the proportion of $(L_t,S)$ in each state. The first two columns of the table correspond one-to-one with Table \ref{tab:Angular-momentum space}. The first eight radial excitations of triply heavy baryons with a given $J^P$ are listed in the row where the symbol $(L_t,S)$ is located in MeV. The numbers under each radial excitation represent the proportion of $(L_t,S)$ corresponding to the row where the number is located. The maximum number in each column has been bolded. \label{tab:ccc bbb Mass spectrum}}
        \renewcommand\arraystretch{1.2}
        \begin{tabular}{@{}l@{}}
            \begin{tabular}{cc|rrrrrrrr|}
                \hline\hline
                \multicolumn{2}{c}{}&\multicolumn{8}{c}{$\Omega_{ccc}$}\\
                \hline
                $J^P$&$(L_t,S)$&5354&5370&5701&5719&5784&5851&5885&6011\\
                \hline
                \multirow{4}{*}{$\dfrac{1}{2}^+$}
                &$(0,\frac{1}{2})$&\textbf{97.34}&2.59&\textbf{98.11}&1.80&\textbf{99.43}&0.55&0.03&\textbf{98.55}\\
                &$(2,\frac{3}{2})$&2.66&\textbf{97.38}&1.86&\textbf{98.09}&0.14&3.58&\textbf{96.35}&1.38\\
                &$(1,\frac{1}{2})$&0.00&0.02&0.04&0.10&0.43&\textbf{95.87}&3.61&0.06\\
                &$(1,\frac{3}{2})$&0.00&0.01&0.00&0.00&0.00&0.00&0.01&0.00\\
            \end{tabular}\\
            \begin{tabular}{cc|rrrrrrrr|}
                \hline
                $J^P$&$(L_t,S)$&4793&5269&5372&5409&5650&5722&5740&5796\\
                \hline
                \multirow{6.5}{*}{$\dfrac{3}{2}^+$}
                &$(0,\frac{3}{2})$&\textbf{99.78}&\textbf{99.79}&0.06&0.03&\textbf{99.71}&0.11&0.05&0.02\\
                &$(2,\frac{3}{2})$&0.12&0.13&\textbf{97.41}&2.48&0.19&\textbf{96.44}&3.17&0.31\\
                &$(2,\frac{1}{2})$&0.09&0.08&2.51&\textbf{97.47}&0.08&3.39&\textbf{96.71}&\textbf{99.31}\\
                &$(1,\frac{1}{2})$&0.00&0.00&0.01&0.00&0.02&0.04&0.05&0.35\\
                &$(1,\frac{3}{2})$&0.00&0.00&0.01&0.00&0.00&0.01&0.01&0.01\\
                &$(3,\frac{3}{2})$&0.00&0.00&0.01&0.01&0.00&0.01&0.01&0.01\\
                \hline
            \end{tabular}\\
            \begin{tabular}{cc|rrrrrrrr|}
            $J^P$&$(L_t,S)$&5374&5409&5725&5747&5801&5852&5883&5887\\
            \hline
            \multirow{5.5}{*}{$\dfrac{5}{2}^+$}
            &$(2,\frac{3}{2})$&\textbf{90.14}&9.82&\textbf{87.88}&11.53&0.54&0.90&48.87&\textbf{50.24}\\
            &$(2,\frac{1}{2})$&9.83&\textbf{90.17}&12.09&\textbf{88.45}&\textbf{99.42}&\textbf{98.99}&0.78&0.25\\
            &$(1,\frac{3}{2})$&0.00&0.00&0.00&0.00&0.00&0.00&0.00&0.00\\
            &$(3,\frac{3}{2})$&0.01&0.00&0.01&0.01&0.01&0.00&0.00&0.01\\
            &$(3,\frac{1}{2})$&0.02&0.00&0.02&0.01&0.02&0.10&\textbf{50.35}&49.50\\
            \hline
            \end{tabular}\\
            \begin{tabular}{cc|rrrrrrrr|}
                $J^P$&$(L_t,S)$&5378&5730&5879&5891&6041&6131&6186&6200\\
                \hline
                \multirow{3}{*}{$\dfrac{7}{2}^+$}
                &$(2,\frac{3}{2})$&\textbf{99.99}&\textbf{99.96}&\textbf{59.87}&40.14&\textbf{99.93}&\textbf{99.83}&9.28&\textbf{85.18}\\
                &$(3,\frac{3}{2})$&0.01&0.00&0.00&0.01&0.00&0.01&0.47&0.50\\
                &$(3,\frac{1}{2})$&0.01&0.03&40.13&\textbf{59.85}&0.06&0.16&\textbf{90.25}&14.32\\
            \end{tabular}\\
            \begin{tabular}{cc|rrrrrrrr|}
            \hline
            $J^P$&$(L_t,S)$&5121&5516&5552&5644&5854&5872&5958&5983\\
            \hline
            \multirow{3}{*}{$\dfrac{1}{2}^-\!$}
            &$(1,\frac{1}{2})$&\textbf{99.93}&\textbf{99.76}&0.30&\textbf{99.87}&\textbf{98.93}&1.10&\textbf{99.93}&\textbf{99.78}\\&$(1,\frac{3}{2})$&0.04&0.22&\textbf{99.67}&0.12&1.05&\textbf{98.86}&0.05&0.16\\
            &$(2,\frac{3}{2})$&0.02&0.02&0.03&0.01&0.02&0.03&0.02&0.06\\
            \end{tabular}\\
            \begin{tabular}{cc|rrrrrrrr|}
            \hline
            $J^P$&$(L_t,S)$&5136&5527&5548&5629&5640&5683&5862&5868\\
            \hline
            \multirow{5}{*}{$\dfrac{3}{2}^-\!$}
            &$(1,\frac{1}{2})$&\textbf{99.90}&\textbf{99.00}&1.21&5.11&\textbf{93.61}&0.98&\textbf{87.24}&12.79\\
            &$(1,\frac{3}{2})$&0.00&0.87&\textbf{98.67}&0.01&0.34&0.05&12.58&\textbf{87.09}\\
            &$(3,\frac{3}{2})$&0.08&0.08&0.02&\textbf{93.12}&5.58&1.27&0.09&0.02\\
            &$(2,\frac{1}{2})$&0.00&0.03&0.07&1.76&0.46&\textbf{97.69}&0.07&0.07\\
            &$(2,\frac{3}{2})$&0.02&0.01&0.03&0.00&0.01&0.01&0.02&0.03\\
            \hline
            \end{tabular}\\
            \begin{tabular}{cc|rrrrrrrr|}
                $J^P$&$(L_t,S)$&5547&5592&5632&5674&5868&5910&5930&5997\\
                \hline
                \multirow{5}{*}{$\dfrac{5}{2}^-\!$}
                &$(1,\frac{3}{2})$&\textbf{99.80}&0.01&0.05&0.05&\textbf{99.77}&0.01&0.09&0.04\\
                &$(3,\frac{1}{2})$&0.06&\textbf{99.27}&0.51&0.23&0.06&\textbf{98.09}&2.32&\textbf{49.69}\\
                &$(3,\frac{3}{2})$&0.07&0.40&\textbf{97.69}&1.84&0.10&1.67&\textbf{97.34}&20.95\\
                &$(2,\frac{1}{2})$&0.06&0.32&1.74&\textbf{97.87}&0.06&0.22&0.24&29.29\\
                &$(2,\frac{3}{2})$&0.01&0.01&0.00&0.01&0.01&0.01&0.02&0.04\\
                \hline\hline
            \end{tabular}
        \end{tabular}%
        \begin{tabular}{@{}l@{}}
            \begin{tabular}{rrrrrrrr}
                \hline\hline
                \multicolumn{8}{c}{$\Omega_{bbb}$}\\
                \hline
                14882&14898&15170&15190&15257&15324&15349&15424\\
                \hline
                \textbf{99.83}&0.17&\textbf{99.90}&0.09&\textbf{99.99}&0.01&0.01&\textbf{99.92}\\
                0.17&\textbf{99.83}&0.10&\textbf{99.90}&0.01&0.08&\textbf{99.91}&0.07\\
                0.00&0.00&0.00&0.01&0.01&\textbf{99.91}&0.08&0.00\\
                0.00&0.00&0.00&0.00&0.00&0.00&0.00&0.00\\
            \end{tabular}\\
            \begin{tabular}{rrrrrrrr}
                \hline
                14365&14792&14901&14937&15117&15192&15209&15266\\
                \hline
                \textbf{99.98}&\textbf{99.98}&0.00&0.00&\textbf{99.98}&0.01&0.00&0.00\\
                0.01&0.01&\textbf{99.93}&0.06&0.01&\textbf{99.94}&0.04&0.01\\
                0.01&0.01&0.06&\textbf{99.94}&0.01&0.04&\textbf{99.96}&\textbf{99.98}\\
                0.00&0.00&0.00&0.00&0.00&0.00&0.00&0.01\\
                0.00&0.00&0.00&0.00&0.00&0.00&0.00&0.00\\
                0.00&0.00&0.00&0.00&0.00&0.00&0.00&0.00\\
                \hline
            \end{tabular}\\
            \begin{tabular}{rrrrrrrr}
                14903&14937&15194&15211&15267&15327&15351&15354\\
                \hline
                \textbf{99.66}&0.33&\textbf{99.53}&0.43&0.03&0.05&\textbf{99.87}&0.08\\
                0.34&\textbf{99.67}&0.46&\textbf{99.56}&\textbf{99.97}&\textbf{99.95}&0.05&0.00\\
                0.00&0.00&0.00&0.00&0.00&0.00&0.00&0.00\\
                0.00&0.00&0.00&0.00&0.00&0.00&0.00&0.00\\
                0.00&0.00&0.00&0.00&0.00&0.00&0.08&\textbf{99.91}\\
                \hline
            \end{tabular}\\
            \begin{tabular}{rrrrrrrr}
                14905&15196&15352&15356&15448&15538&15600&15608\\
                \hline
                \textbf{100.0}&\textbf{100.0}&\textbf{95.09}&4.91&\textbf{100.0}&\textbf{100.0}&0.67&\textbf{99.28}\\
                0.00&0.00&0.00&0.00&0.00&0.00&0.03&0.01\\
                0.00&0.00&4.91&\textbf{95.09}&0.00&0.00&\textbf{99.29}&0.71\\
            \end{tabular}\\
            \begin{tabular}{rrrrrrrr}
                \hline
                14687&15021&15041&15143&15299&15304&15398&15422\\
                \hline
                \textbf{99.99}&\textbf{99.94}&0.06&\textbf{99.99}&\textbf{99.03}&0.97&\textbf{99.99}&\textbf{99.99}\\
                0.00&0.06&\textbf{99.93}&0.01&0.97&\textbf{99.02}&0.00&0.01\\
                0.00&0.00&0.00&0.00&0.00&0.00&0.00&0.01\\
            \end{tabular}\\
            \begin{tabular}{rrrrrrrr}
                \hline
                14692&15025&15041&15123&15143&15180&15302&15304\\
                \hline
                \textbf{99.99}&\textbf{99.92}&0.08&0.05&\textbf{99.92}&0.02&\textbf{96.91}&3.09\\
                0.00&0.07&\textbf{99.90}&0.00&0.01&0.01&3.08&\textbf{96.90}\\
                0.01&0.01&0.00&\textbf{99.90}&0.05&0.04&0.01&0.00\\
                0.00&0.00&0.01&0.04&0.02&\textbf{99.92}&0.00&0.01\\
                0.00&0.00&0.00&0.00&0.00&0.00&0.00&0.00\\
                \hline
            \end{tabular}\\
            \begin{tabular}{rrrrrrrr}
                15042&15093&15124&15180&15305&15352&15364&15436\\
                \hline
                \textbf{99.99}&0.00&0.00&0.00&\textbf{99.98}&0.00&0.00&0.00\\
                0.01&\textbf{99.99}&0.00&0.01&0.00&\textbf{99.96}&0.07&\textbf{95.30}\\
                0.01&0.00&\textbf{99.94}&0.05&0.01&0.03&\textbf{99.92}&4.50\\
                0.00&0.01&0.05&\textbf{99.93}&0.00&0.01&0.01&0.20\\
                0.00&0.00&0.00&0.00&0.00&0.00&0.00&0.00\\
                \hline\hline
            \end{tabular}
        \end{tabular}%
    \end{table*}
    \endgroup
    
    \begingroup
    \squeezetable
    \begin{table*}[htbp]
        \caption{
        The mass spectra of the $\Omega_{bcc}$ and $\Omega_{bbc}$ baryons and the proportion of $(l,L,L_t,S)$ in each state.
            \label{tab:bcc bbc Mass spectrum}}
        \renewcommand\arraystretch{1.2}
        \begin{tabular}{@{}l@{}}
            \begin{tabular}{cc|rrrrrrrr|}
                \hline\hline
                \multicolumn{2}{c}{}&\multicolumn{8}{c}{$\Omega_{bcc}$}\\
                \hline
                $J^P$&$(l,L,L_t,S)$&8017&8463&8546&8564&8605&8657&8667&8825\\
                \hline
                \multirow{7}{*}{$\dfrac{1}{2}^+$}
                &$(0,0,0,\frac{1}{2})$&\textbf{99.91}&\textbf{99.91}&0.01&0.20&\textbf{99.32}&0.18&0.29&\textbf{99.88}\\
                &$(1,1,0,\frac{1}{2})$&0.01&0.01&0.38&\textbf{99.40}&0.20&0.01&0.00&0.01\\
                &$(0,2,2,\frac{3}{2})$&0.00&0.00&\textbf{98.56}&0.37&0.00&0.07&1.41&0.00\\
                &$(2,0,2,\frac{3}{2})$&0.07&0.06&1.00&0.02&0.35&0.51&\textbf{97.76}&0.06\\
                &$(1,1,1,\frac{1}{2})$&0.01&0.02&0.03&0.01&0.12&\textbf{99.22}&0.53&0.05\\
                &$(2,2,1,\frac{1}{2})$&0.00&0.00&0.02&0.00&0.00&0.00&0.00&0.00\\
                &$(2,2,1,\frac{3}{2})$&0.00&0.00&0.00&0.01&0.00&0.00&0.00&0.00\\
            \end{tabular}\\
            \begin{tabular}{cc|rrrrrrrr|}
                \hline
                $J^P$&$(l,L,L_t,S)$&8030&8469&8545&8549&8603&8622&8660&8670\\
                \hline
                \multirow{11}{*}{$\dfrac{3}{2}^+$}
                &$(0,0,0,\frac{3}{2})$&\textbf{99.70}&\textbf{99.38}&0.05&0.02&\textbf{98.35}&0.08&0.18&0.08\\
                &$(2,2,0,\frac{3}{2})$&0.20&0.53&0.00&0.01&1.19&0.00&0.00&0.00\\
                &$(0,2,2,\frac{3}{2})$&0.01&0.00&\textbf{53.37}&44.20&0.08&0.95&0.06&0.67\\
                &$(2,0,2,\frac{3}{2})$&0.04&0.04&0.67&0.66&0.03&0.07&8.31&39.60\\
                &$(0,2,2,\frac{1}{2})$&0.00&0.00&43.57&\textbf{53.77}&0.00&0.60&0.05&1.45\\
                &$(1,1,2,\frac{1}{2})$&0.00&0.00&1.50&0.00&0.06&\textbf{98.29}&0.01&0.10\\
                &$(2,0,2,\frac{1}{2})$&0.03&0.03&0.79&1.18&0.15&0.00&2.13&\textbf{57.54}\\
                &$(1,1,1,\frac{1}{2})$&0.01&0.02&0.01&0.15&0.12&0.00&\textbf{89.25}&0.55\\
                &$(2,2,1,\frac{1}{2})$&0.00&0.00&0.01&0.00&0.00&0.00&0.00&0.00\\
                &$(2,2,1,\frac{3}{2})$&0.00&0.00&0.01&0.01&0.02&0.00&0.01&0.01\\
                &$(2,2,3,\frac{3}{2})$&0.00&0.00&0.02&0.00&0.00&0.00&0.00&0.00\\
                \hline
            \end{tabular}\\
            \begin{tabular}{cc|rrrrrrrr|}
                $J^P$&$(l,L,L_t,S)$&8548&8550&8625&8668&8679&8862&8863&8957\\
                \hline
                \multirow{10}{*}{$\dfrac{5}{2}^+$}
                &$(0,2,2,\frac{3}{2})$&18.23&\textbf{78.37}&1.68&0.57&1.85&28.19&\textbf{70.13}&1.12\\
                &$(2,0,2,\frac{3}{2})$&0.34&1.52&0.20&19.41&\textbf{77.78}&0.24&0.59&0.44\\
                &$(2,0,2,\frac{1}{2})$&2.02&0.48&0.37&\textbf{77.22}&19.02&0.82&0.32&0.47\\
                &$(1,1,2,\frac{1}{2})$&1.65&0.55&\textbf{96.86}&0.15&0.79&0.96&0.20&\textbf{97.25}\\
                &$(0,2,2,\frac{1}{2})$&\textbf{77.74}&19.06&0.88&2.64&0.56&\textbf{69.75}&28.71&0.69\\
                &$(2,2,1,\frac{3}{2})$&0.00&0.00&0.00&0.00&0.01&0.01&0.01&0.00\\
                &$(2,2,3,\frac{3}{2})$&0.01&0.00&0.00&0.00&0.00&0.01&0.00&0.01\\
                &$(1,3,3,\frac{1}{2})$&0.00&0.01&0.00&0.00&0.00&0.02&0.04&0.00\\
                &$(2,2,3,\frac{1}{2})$&0.00&0.01&0.00&0.00&0.00&0.00&0.01&0.01\\
                &$(3,1,3,\frac{1}{2})$&0.00&0.00&0.00&0.00&0.00&0.00&0.00&0.00\\
                \hline
            \end{tabular}\\
            \begin{tabular}{cc|rrrrrrrr|}
                $J^P$&$(l,L,L_t,S)$&8552&8676&8866&8985&9029&9033&9080&9087\\
                \hline
                \multirow{6}{*}{$\dfrac{7}{2}^+$}
                &$(0,2,2,\frac{3}{2})$&\textbf{97.70}&3.16&\textbf{98.78}&\textbf{73.97}&22.35&4.45&0.02&0.03\\
                &$(2,0,2,\frac{3}{2})$&2.28&\textbf{96.84}&1.15&25.52&\textbf{68.46}&5.22&1.05&0.99\\
                &$(2,2,3,\frac{3}{2})$&0.00&0.00&0.01&0.00&0.04&1.82&30.45&\textbf{64.35}\\
                &$(1,3,3,\frac{1}{2})$&0.01&0.00&0.06&0.49&8.96&\textbf{87.47}&0.02&2.62\\
                &$(2,2,3,\frac{1}{2})$&0.00&0.00&0.00&0.00&0.07&0.47&\textbf{68.46}&30.35\\
                &$(3,1,3,\frac{1}{2})$&0.00&0.00&0.00&0.00&0.11&0.58&0.00&1.66\\
            \end{tabular}\\
            \begin{tabular}{cc|rrrrrrrr|}
                \hline
                $J^P$&$(l,L,L_t,S)$&8310&8319&8389&8668&8675&8771&8809&8813\\
                \hline
                \multirow{4}{*}{$\dfrac{1}{2}^-\!$}
                &$(0,1,1,\frac{1}{2})$&20.76&\textbf{79.21}&0.04&22.25&\textbf{77.66}&0.15&\textbf{66.10}&33.28\\
                &$(1,0,1,\frac{1}{2})$&0.02&0.01&\textbf{99.92}&0.09&0.03&\textbf{99.67}&0.89&0.01\\
                &$(0,1,1,\frac{3}{2})$&\textbf{79.20}&20.77&0.04&\textbf{77.63}&22.31&0.15&32.93&\textbf{66.72}\\
                &$(2,1,2,\frac{3}{2})$&0.03&0.01&0.00&0.03&0.01&0.03&0.09&0.00\\
            \end{tabular}\\
            \begin{tabular}{cc|rrrrrrrr|}
                \hline
                $J^P$&$(l,L,L_t,S)$&8322&8323&8397&8677&8678&8746&8777&8808\\
                \hline
                \multirow{8}{*}{$\dfrac{3}{2}^-\!$}
                &$(0,1,1,\frac{1}{2})$&\textbf{85.60}&14.04&0.29&\textbf{94.30}&5.51&0.00&0.12&\textbf{96.51}\\
                &$(1,0,1,\frac{1}{2})$&0.54&0.13&\textbf{99.33}&0.17&0.13&0.00&\textbf{99.68}&0.13\\
                &$(0,1,1,\frac{3}{2})$&13.81&\textbf{85.78}&0.37&5.46&\textbf{94.31}&0.00&0.18&2.78\\
                &$(0,3,3,\frac{3}{2})$&0.00&0.00&0.00&0.00&0.00&\textbf{98.19}&0.00&0.00\\
                &$(2,1,3,\frac{3}{2})$&0.02&0.03&0.00&0.03&0.02&1.51&0.00&0.17\\
                &$(1,2,2,\frac{1}{2})$&0.00&0.00&0.00&0.03&0.01&0.28&0.00&0.39\\
                &$(2,1,2,\frac{1}{2})$&0.00&0.02&0.00&0.00&0.02&0.00&0.01&0.00\\
                &$(2,1,2,\frac{3}{2})$&0.02&0.00&0.00&0.02&0.01&0.01&0.01&0.03\\
                \hline
            \end{tabular}\\
            \begin{tabular}{cc|rrrrrrrr|}
                $J^P$&$(l,L,L_t,S)$&8330&8684&8743&8747&8811&8818&8856&8863\\
                \hline
                \multirow{10}{*}{$\dfrac{5}{2}^-\!$}
                &$(0,1,1,\frac{3}{2})$&\textbf{99.93}&\textbf{99.89}&0.01&0.01&\textbf{99.08}&0.26&0.58&0.00\\
                &$(0,3,3,\frac{1}{2})$&0.00&0.00&45.18&\textbf{51.69}&0.00&1.08&0.21&1.94\\
                &$(1,2,3,\frac{1}{2})$&0.00&0.00&2.24&0.01&0.22&\textbf{88.07}&0.01&2.20\\
                &$(2,1,3,\frac{1}{2})$&0.03&0.03&0.95&1.09&0.09&0.53&12.64&\textbf{61.73}\\
                &$(3,0,3,\frac{1}{2})$&0.00&0.00&0.03&0.00&0.01&7.16&0.04&0.02\\
                &$(0,3,3,\frac{3}{2})$&0.00&0.00&\textbf{50.65}&46.00&0.03&1.71&0.07&0.56\\
                &$(2,1,3,\frac{3}{2})$&0.02&0.02&0.92&0.81&0.10&1.19&6.08&30.67\\
                &$(1,2,2,\frac{1}{2})$&0.01&0.05&0.00&0.38&0.44&0.00&\textbf{79.06}&2.54\\
                &$(2,1,2,\frac{1}{2})$&0.00&0.00&0.00&0.00&0.01&0.00&0.22&0.33\\
                &$(2,1,2,\frac{3}{2})$&0.01&0.01&0.01&0.01&0.02&0.00&1.08&0.01\\
                \hline\hline
            \end{tabular}
        \end{tabular}%
        \begin{tabular}{@{}l@{}}
            \begin{tabular}{rrrrrrrr}
                \hline\hline
                \multicolumn{8}{c}{$\Omega_{bbc}$}\\
                \hline
                11204&11621&11702&11746&11774&11832&11837&11948\\
                \hline
                \textbf{99.98}&\textbf{99.95}&0.07&\textbf{99.16}&0.78&0.04&0.01&\textbf{99.83}\\
                0.01&0.03&0.03&0.74&\textbf{99.11}&0.06&0.07&0.06\\
                0.00&0.00&0.27&0.00&0.07&26.76&\textbf{72.99}&0.00\\
                0.01&0.02&\textbf{99.57}&0.07&0.02&0.04&0.21&0.09\\
                0.00&0.00&0.05&0.02&0.02&\textbf{73.09}&26.72&0.01\\
                0.00&0.00&0.00&0.00&0.00&0.00&0.00&0.00\\
                0.00&0.00&0.00&0.00&0.00&0.00&0.00&0.00\\
            \end{tabular}\\
            \begin{tabular}{rrrrrrrr}
                \hline
                11221&11627&11700&11708&11752&11813&11827&11837\\
                \hline
                \textbf{99.89}&\textbf{99.28}&0.07&0.01&\textbf{99.76}&0.06&0.00&0.04\\
                0.09&0.69&0.00&0.00&0.07&0.00&0.00&0.00\\
                0.01&0.01&0.07&0.39&0.00&0.36&21.69&9.58\\
                0.01&0.01&32.86&\textbf{66.58}&0.06&0.03&0.15&0.01\\
                0.00&0.00&0.44&0.44&0.00&0.21&\textbf{77.54}&2.74\\
                0.00&0.00&0.12&0.00&0.05&\textbf{99.13}&0.01&0.02\\
                0.00&0.01&\textbf{66.43}&32.57&0.02&0.08&0.61&0.01\\
                0.00&0.00&0.01&0.00&0.02&0.13&0.00&\textbf{87.59}\\
                0.00&0.00&0.00&0.00&0.00&0.00&0.00&0.00\\
                0.00&0.00&0.00&0.00&0.01&0.00&0.00&0.01\\
                0.00&0.00&0.00&0.00&0.00&0.00&0.00&0.00\\
                \hline
            \end{tabular}\\
            \begin{tabular}{rrrrrrrr}
                11699&11712&11812&11830&11846&11986&11996&12091\\
                \hline
                0.03&0.70&0.69&\textbf{54.28}&44.48&0.06&0.74&0.03\\
                13.23&\textbf{86.04}&0.01&0.40&0.13&13.60&\textbf{85.60}&0.00\\
                \textbf{86.16}&13.05&0.08&0.32&0.18&\textbf{85.67}&13.46&0.03\\
                0.04&0.01&\textbf{98.73}&1.22&0.00&0.03&0.01&\textbf{99.88}\\
                0.54&0.21&0.49&43.77&\textbf{55.21}&0.64&0.19&0.03\\
                0.00&0.00&0.00&0.00&0.00&0.00&0.00&0.00\\
                0.00&0.00&0.00&0.00&0.00&0.00&0.00&0.00\\
                0.00&0.00&0.00&0.00&0.00&0.00&0.00&0.00\\
                0.00&0.00&0.00&0.00&0.00&0.00&0.00&0.00\\
                0.00&0.00&0.00&0.00&0.00&0.00&0.00&0.02\\
                \hline
            \end{tabular}\\
            \begin{tabular}{rrrrrrrr}
                11715&11833&11998&12160&12174&12189&12198&12236\\
                \hline
                0.97&\textbf{99.25}&0.97&\textbf{97.20}&0.05&25.72&\textbf{76.22}&0.00\\
                \textbf{99.03}&0.74&\textbf{99.03}&2.80&0.65&\textbf{73.50}&23.67&0.04\\
                0.00&0.00&0.00&0.00&0.04&0.11&0.03&\textbf{59.84}\\
                0.00&0.00&0.00&0.00&0.19&0.00&0.00&0.10\\
                0.00&0.00&0.00&0.00&0.01&0.02&0.01&39.98\\
                0.00&0.00&0.00&0.00&\textbf{99.05}&0.65&0.06&0.04\\
            \end{tabular}\\
            \begin{tabular}{rrrrrrrr}
                \hline
                11496&11559&11571&11823&11915&11919&11971&11977\\
                \hline
                0.23&9.02&\textbf{90.75}&0.08&10.65&\textbf{89.28}&12.02&\textbf{87.47}\\
                \textbf{99.41}&0.52&0.07&\textbf{99.81}&0.15&0.03&0.90&0.16\\
                0.36&\textbf{90.45}&9.18&0.11&\textbf{89.20}&10.69&\textbf{87.07}&12.37\\
                0.00&0.00&0.00&0.00&0.00&0.00&0.01&0.00\\
            \end{tabular}\\
            \begin{tabular}{rrrrrrrr}
                \hline
                11506&11576&11578&11830&11922&11928&11980&11981\\
                \hline
                0.06&33.20&\textbf{66.73}&0.02&\textbf{55.74}&44.22&13.84&35.70\\
                \textbf{99.93}&0.02&0.04&\textbf{99.97}&0.01&0.01&0.11&0.11\\
                0.00&\textbf{66.78}&33.22&0.00&44.23&\textbf{55.76}&17.16&33.35\\
                0.00&0.00&0.00&0.00&0.00&0.00&0.56&0.28\\
                0.00&0.00&0.00&0.00&0.00&0.00&\textbf{67.80}&30.20\\
                0.00&0.00&0.00&0.00&0.00&0.00&0.00&0.01\\
                0.00&0.00&0.00&0.00&0.00&0.00&0.41&0.34\\
                0.00&0.00&0.00&0.00&0.02&0.00&0.13&0.01\\
                \hline
            \end{tabular}\\
            \begin{tabular}{rrrrrrrr}
                11584&11878&11927&11975&11982&11987&12014&12018\\
                \hline
                \textbf{99.99}&0.01&\textbf{99.95}&0.00&0.69&\textbf{99.17}&0.16&0.01\\
                0.00&0.00&0.00&1.25&0.05&0.00&0.00&0.03\\
                0.00&0.90&0.00&0.00&0.05&0.00&0.14&0.00\\
                0.00&0.09&0.00&\textbf{88.82}&9.09&0.07&0.07&0.53\\
                0.00&\textbf{98.96}&0.01&0.07&0.05&0.00&0.02&0.00\\
                0.00&0.00&0.00&0.18&0.80&0.01&0.01&0.07\\
                0.00&0.01&0.01&9.47&\textbf{87.77}&0.62&0.47&0.60\\
                0.00&0.01&0.00&0.00&0.00&0.00&0.01&0.03\\
                0.00&0.00&0.01&0.18&1.43&0.00&20.73&\textbf{77.60}\\
                0.00&0.02&0.03&0.04&0.05&0.12&\textbf{78.39}&21.12\\
                \hline\hline
            \end{tabular}
        \end{tabular}%
    \end{table*}
    \endgroup
    
\subsubsection{$\Omega_{ccc}$ and $\Omega_{bbb}$ baryons}
The upper panel of Fig.~\ref{fig:Mass spectrum} displays our predicted mass spectra for $\Omega_{ccc}$ and $\Omega_{bbb}$ baryons with quantum numbers $J^P=\frac{1}{2}^\pm$, $\frac{3}{2}^\pm$, $\frac{5}{2}^\pm$, and $\frac{7}{2}^+$. For these two kinds of baryons, our predictions for the lowest $\frac{3}{2}^+$, $\frac{1}{2}^-$ and $\frac{3}{2}^-$ states agree well with the latest lattice results \cite{Dhindsa:2024erk,Brown:2014ena}, though other states show significant deviations from the only available lattice data \cite{Padmanath:2013zfa,Meinel:2012qz}. 
We note that the highly excited states in Ref.~\cite{Padmanath:2013zfa} carry substantial uncertainties, limiting their reliability as benchmarks. Despite these quantitative differences, both approaches predict qualitatively similar energy level structures. 

In our model, $\Omega_{ccc}$ and $\Omega_{bbb}$ differ only in quark mass, resulting in nearly identical energy level structures (upper panel of Fig.~\ref{fig:Mass spectrum}). Table~\ref{tab:ccc bbb Mass spectrum} reveals that most states are dominated by single total orbital angular momentum excitations ($L_t$), with minimal mixing between different $L_t$ values. We therefore characterize states by their dominant orbital component ($S$, $P$, $D$):

\begin{itemize}
    \item $\frac{1}{2}^+$ system: Contains two nearly degenerate $S$-$D$ wave pairs, distinct from simple spin excitation pairs
    \item $\frac{3}{2}^+$ system: Features the ground states ($1S$), followed by $2S$-$3S$ states and two $1D$-$2D$ spin excitation pairs
    \item $\frac{5}{2}^+$ system: Shows two $1D$-$2D$ pairs with moderate spin $S= 1/2$ and $S= 3/2$ state mixing
    \item $\frac{7}{2}^+$ system: Contains only $D$-wave states ($S=3/2$ required for $J=7/2$ coupling)
    \item Negative parity systems: Exhibit clean $P$-wave ($1P$, $2P$, $3P$) and $F$-wave patterns. A particularly interesting state appears in the $\frac{5}{2}^-$ system, where $L_t=2$ emerges from coupled orbital angular momenta (1 and 2) as detailed in Table~\ref{tab:Angular-momentum space}.
\end{itemize}

We clarify a common misconception regarding $J^P=\frac{1}{2}^+$ states in three-identical-particle systems. Contrary to claims in Refs.~\cite{Yang:2019lsg,Yu:2025gdg}, $S$-wave bound states are indeed possible despite the Pauli principle. While no symmetric $S=1/2$ spin state exists for three identical fermions, Eq.~\eqref{eq:Phi1} demonstrates how symmetric spatial states can be constructed through appropriate combinations of two-particle symmetric states. This explains the presence of closely spaced $S$-wave and $D$-wave states in the $\frac{1}{2}^+$ channel of $\Omega_{ccc}$ and $\Omega_{bbb}$ mass spectra.

It is true that the $J^P=\frac{1}{2}^+$ $S$-wave states are also calculated by using harmonic oscillator expansion in Ref.~\cite{Liu:2019vtx}. However, for the mass spectra of $\Omega_{ccc}$ and $\Omega_{bbb}$ baryons calculated in Ref.~\cite{Liu:2019vtx}, the mass of $D$-wave $\Omega_{ccc}$ baryon with $J^P=\frac{1}{2}^+$ is higher than that of $S$-wave $\Omega_{ccc}$ baryon, while for the $\Omega_{bbb}$ baryons, it is the opposite. Our calculation results show that the energy level order of $\Omega_{ccc}$ and $\Omega_{bbb}$ is exactly the same. As shown in Fig.~\ref{fig:Mass spectrum}, the $\Omega_{ccc}$ and $\Omega_{bbb}$ baryons exhibit similar mass spectra, reflected in their comparable excitation energies. Since the kinetic energy and spin-dependent terms contain constituent quark masses, there are small differences in the excitation energies between $\Omega_{ccc}$ and $\Omega_{bbb}$. Based on the heavy-quark flavor symmetry, this similarity in the mass spectra may be a characteristic of triply heavy baryons.

\subsubsection{$\Omega_{bcc}$ and $\Omega_{bbc}$ baryons}
The lower panel of Fig.~\ref{fig:Mass spectrum} presents our predicted mass spectra for $\Omega_{bcc}$ and $\Omega_{bbc}$ baryons across various quantum states ($J^P=\frac{1}{2}^\pm$, $\frac{3}{2}^\pm$, $\frac{5}{2}^\pm$, $\frac{7}{2}^+$). Our calculations yield masses for the lowest $\frac{1}{2}^+$ and $\frac{3}{2}^+$ states that are slightly elevated (by 10--15 MeV) compared to lattice QCD results \cite{Mathur:2018epb,Brown:2014ena,Burch:2015pka}, though still consistent within theoretical uncertainties (see Table~\ref{tab:Lattice results}).

While $\Omega_{bcc}$ and $\Omega_{bbc}$ share similar symmetry properties, their distinct quark compositions (``one heavy + two light" versus ``two heavy + one light") lead to characteristic spectral differences:

\begin{itemize}
    \item $\frac{1}{2}^+$ system: Contains the ground states ($1S$), followed by $2S$, $3S$, and four distinct $D$-wave excitations. The state ordering differs between $\Omega_{bcc}$ and $\Omega_{bbc}$ due to mass effects (Table~\ref{tab:bcc bbc Mass spectrum}).
    
    \item $\frac{3}{2}^+$ system: Features $1S$/$2S$ states (shifted $\sim$20 MeV higher than $\frac{1}{2}^+$), followed by mode-specific $1D$ excitations:
    \begin{itemize}
        \item $\Omega_{bcc}$: $(0,2,2)$-mode doublet
        \item $\Omega_{bbc}$: $(2,0,2)$-mode doublet
    \end{itemize}
    Higher states include $3S$ and three $2D$ configurations.
    
    \item $\frac{5}{2}^+$ system: Dominated by $1D$ spin doublets with:
    \begin{itemize}
        \item $\Omega_{bcc}$: $(0,2,2)$-mode doublet
        \item $\Omega_{bbc}$: $(2,0,2)$-mode doublet
    \end{itemize}
    plus three higher $2D$ states.
    
    \item $\frac{7}{2}^+$ system: Contains only $S=3/2$ $D$-wave states:
    \begin{itemize}
        \item $\Omega_{bcc}$: $(0,2,2)$-$1D$ and $(2,0,2)$-$2D$
        \item $\Omega_{bbc}$: $(2,0,2)$-$1D$ and $(0,2,2)$-$2D$
    \end{itemize}
    Here, we can clearly see the mixing of $(0,2,2)$-mode and $(2,0,2)$-mode, as predicted before. 
    
    \item $\frac{1}{2}^-$ system: Shows clear $P$-wave structures with mode-dependent ordering

    \item $\frac{3}{2}^-$ system: Similar to $\frac{1}{2}^-$ but with $\sim$10 MeV mass shifts

    \item $\frac{5}{2}^-$ system: $\Omega_{bcc}$ exhibits two $P$-wave states while $\Omega_{bbc}$ shows a $P$-wave state and an $F$-wave state
\end{itemize}

Prior research primarily considered only $\lambda$-mode excitation \cite{Luo:2023sne,Peng:2024pyl}, or posited that mode excitation is dominated by heavy quarks \cite{Yu:2025gdg}. However, we propose that all distinct mode excitations coexist. Within the established theoretical framework, it is not feasible to selectively exclude any specific mode. These systematic spectral patterns, detailed in Table~\ref{tab:bcc bbc Mass spectrum}, arise from the interplay between quark masses and QCD dynamics.

    \begin{table}[htbp]
        \caption{Comparison of our results with other studies, including the latest lattice QCD results [4793(5)(7) MeV and 5094(12)(13) MeV for $\frac{3}{2}^+$ and $\frac{3}{2}^-$ states] from Ref.~\cite{Dhindsa:2024erk}.\label{tab:Lattice results}}
        \begin{ruledtabular}
            \begin{tabular}{cclcl}
                &\multicolumn{2}{c}{$\Omega_{ccc}$}&\multicolumn{2}{c}{$\Omega_{bbb}$}\\
                \cline{2-3}\cline{4-5}
                $J^P$&Our&Lattice~\cite{Padmanath:2013zfa}&Our&Lattice~\cite{Meinel:2010pw,Meinel:2012qz}\\
                \hline
                \multirow{6}{*}{$\dfrac{1}{2}^+$}
                &5354&5399(13)&14882&14938(14)(12)\\
                &5370&5405(14)&14898&14953(13)(13)\\
                &5701&6039(33)&15170&\\
                &5719&6083(42)&15190& \\
                &5784&6468(80)&15257& \\
                &5851&6712(56)&15324& \\
                \hline
                \multirow{8}{*}{$\dfrac{3}{2}^+$}
                &4793&4763(6)&14365&14371(4)(11)\\
                &5269&5317(31)&14792&14840(11)(9)\\
                &5372&5430(13)&14901&14958(13)(12)\\
                &5409&5465(13)&14937&15005(14)(13)\\
                &5650&6094(40)&15117& \\
                &5722&6623(46)&15192& \\
                &5740&6641(43)&15209& \\
                &5796&6741(46)&15266& \\
                \hline
                \multirow{2}{*}{$\dfrac{5}{2}^+$}
                &5374&5406(15)&14903&14964(13)(12)\\
                &5409&5464(15)&14937&15007(14)(13)\\
                \hline
                \rule{0pt}{4ex}$\dfrac{7}{2}^+$
                &5378&5397(49)&14905&14969(12)(12)\\[5pt]
                \hline
                \multirow{5}{*}{$\dfrac{1}{2}^-$}
                &5121&5120(9)&14687&14706.3(5.8)(7.4)\\
                &5516&5612(31)&15021& \\
                &5552&5631(43)&15041& \\
                &5644&5749(21)&15143& \\
                &5854&6132(69)&15299& \\
                \hline
                \multirow{6}{*}{$\dfrac{3}{2}^-$}
                &5136&5124(13)&14692&14714.0(5.5)(7.2)\\
                &5527&5662(31)&15025& \\
                &5548&5724(44)&15041& \\
                &5629&5765(34)&15123& \\
                &5640&6299(44)&15143& \\
                &5683&6326(51)&15180& \\
                \hline
                \multirow{4}{*}{$\dfrac{5}{2}^-$}
                &5547&5516(64)&15042& \\
                &5592&5709(25)&15093& \\
                &5632&5710(24)&15124& \\
                &5674&5712(32)&15180& \\
            \end{tabular}
        \end{ruledtabular}
        
        \begin{tabular}{ccccc}
            Hadrons&Our&Lattice~\cite{Mathur:2018epb}&Lattice~\cite{Brown:2014ena}&Lattice~\cite{Burch:2015pka}\\
            \hline
            $\Omega_{bcc}(1/2^+)$&8017&8005(6)(11)&8007(9)(20)&7984(27)(12)\\
            $\Omega_{bcc}(3/2^+)$&8030&8026(7)(11)&8037(9)(20)&8012.8(5.6)(2.2)\\
            $\Omega_{bcc}(3/2^-)$&8322&$-$&$-$&8461.8(41)(34)\\
            \hline
            $\Omega_{bbc}(1/2^+)$&11204&11194(5)(12)&11195(8)(20)&11182(27)(13)\\
            $\Omega_{bbc}(3/2^+)$&11221&11211(6)(12)&11229(8)(20)&11203.8(7.0)(1.7)\\
            $\Omega_{bbc}(1/2^-)$&11496&$-$&$-$&11557(74)(28)\\
            $\Omega_{bbc}(3/2^-)$&11506&$-$&$-$&11587(12)(2)\\
            \hline
            $\Omega_{bbb}(3/2^+)$&14365&$-$&14366(9)(20)&14369(21)(14)\\
            \hline\hline
        \end{tabular}
    \end{table}

    \subsection{Radiative decay}
    Table \ref{tab:Strong decay threshold} presents the strong decay thresholds for triply heavy baryons. Our analysis reveals many states lying below these thresholds, making them prime candidates for radiative transitions. We have systematically calculated the radiative decay properties of $\Omega_{ccc}$, $\Omega_{bbb}$, $\Omega_{bcc}$, and $\Omega_{bbc}$ baryons with masses below 5.486 GeV ($\varXi_{cc}^{++}D^0$ threshold), 15.25 GeV, 8.58 GeV and 11.72 GeV, respectively. 
All considered initial states lie below the strong decay threshold, precluding OZI-allowed two-body strong decays. Consequently, these radiative transitions may serve as crucial signatures for identifying these states in future experiments.

    \begin{table}[htbp]
        \caption{Strong decay thresholds (mass values in MeV) for triply heavy baryons. Masses of $\varXi_{cb}^+$ and $\varXi_{bb}^0$ use latest lattice results: 6945(22)(14) MeV \cite{Mathur:2018epb} and 10143(30)(23) MeV \cite{Brown:2014ena}. \label{tab:Strong decay threshold}}
        \begin{tabular}{c|c|c|c}
            \hline\hline
            $\Omega_{ccc}$&$\Omega_{bcc}$&$\Omega_{bbc}$&$\Omega_{bbb}$\\
            \hline
            $\varXi_{cc}^{++}D^0:5486$&$\varXi_{cb}^+D^0:8810$&$\varXi_{bb}D^0:12008$&$\varXi_{bb}B^-:15422$\\
            &$\varXi_{cc}^{++}B^-:8901$&$\varXi_{cb}^+B^-:12224$&\\
            \hline\hline
        \end{tabular}
    \end{table}
    
Through our calculations and comparison with previous studies, we have identified two fundamental patterns governing radiative decays in our model:
\begin{enumerate}
    \item \textbf{Three identical quarks}: Radiative transitions between states with conserved parity ($P_i = P_f$) and total spin ($S_i = S_f$) typically exhibit larger partial widths.
    
    \item \textbf{Two identical quarks}: The dominant transitions occur between states with conserved total spin ($S_i = S_f$) but flipped parity ($P_i = -P_f$), yielding larger partial widths.
\end{enumerate}
The process of total spin invariance leads to electrical transitions, hence tending to have a larger width. The above two patterns indicate that the influence of identical particle symmetry on baryon radiative decay is significant. For $\Omega_{ccc}$ and $\Omega_{bbb}$ baryons, the radiative decay width from $D$-wave to $S$-wave, $F$-wave to $P$-wave, etc. is larger, while for $\Omega_{bcc}$ and $\Omega_{bbc}$ baryons, the width from $P$-wave to $S$-wave, $D$-wave to $P$-wave, etc. is larger. The radiative decay width is primarily determined by the total orbital angular momenta~$L_t$ of the initial and final states. Therefore, we denote transitions from an~$X$-wave baryon to a~$Y$-wave baryon as~$X \to Y$, where~$X$ and~$Y$ represent the respective orbital angular momentum states of the particles.

    \subsubsection{The $\Omega_{ccc}$ baryons}
    Table \ref{tab:Radiative decay widths of ccc} presents the radiative decay widths for $\Omega_{ccc}$ baryons below the $\varXi_{cc}^{++}D^0$ threshold. Our calculations reveal several key features:

\begin{itemize}
    \item \mbox{$S\to S$ transitions:} The two $S$-wave states $\Omega_{ccc}(5354)1/2^+$ and $\Omega_{ccc}(5269)3/2^+$ decay to the ground state with relatively small decay width 4.99 keV and 0.68 keV.
    
    \item \mbox{$D\to S$ transitions:} Transitions to the ground state $\Omega_{ccc}(4793)3/2^+$ with $S=3/2$ $D$-wave initial states exhibit particularly large partial widths, all in the hundreds of keV. The widths of the two $S=1/2$ $D$-wave states are only 9.96 keV and 24.68 keV due to the absence of electrical transitions.
    
    \item \mbox{$P\to S$ transitions:} The $P$-wave states $\Omega_{ccc}(5121)1/2^-$ and $\Omega_{ccc}(5136)3/2^-$ decay to the ground state with widths of $\mathcal{O}(1)$ keV, which is consistent with Ref.~\cite{Liu:2019vtx}.
    
    \item \mbox{$X\to P$  transitions ($X$ stands for any particle):} Processes with $P$-wave or higher excited final states generally yield smaller widths compared to those with $S$-wave final states.
\end{itemize}
It can be observed that our partial calculations differ significantly from those in Ref.~\cite{Liu:2019vtx}. We attribute this discrepancy to two main factors: (1) Ref.~\cite{Liu:2019vtx} does not account for the symmetry of three identical particles; (2) the potential models employed differ between the studies. Notably, the results obtained in Ref.~\cite{Liu:2019vtx} are consistent with the radiative decay characteristics of two-identical-quark baryons. 
    
    \subsubsection{The $\Omega_{bbb}$ baryons}

    Table \ref{tab:Radiative decay widths of bbb} presents radiative decay widths for $\Omega_{bbb}$ baryons with masses below 15.25 GeV. These widths are generally smaller than those of $\Omega_{ccc}$ baryons due to differences in both quark charge and mass between the $b$ and $c$ quarks. Many transitions are negligible (0.00 keV). The well-resolved mass spectrum of these three-identical-particle systems facilitates comprehensive calculation of radiative widths for nearly all $\Omega_{bbb}$ states shown in Fig.~\ref{fig:Mass spectrum}. Key observations include:

\begin{itemize}
    \item \mbox{$S\to S$ transitions:} The transitions from $S$-wave to $S$-wave are relatively small, only 0.01 keV.
    
    \item \mbox{$D\to S$ transitions:} Similar to $\Omega_{ccc}$ cases, the larger partial widths occur for $D$-wave to $S$-wave transitions with conserved total spin, while the processes of total spin change are two orders of magnitude less.

    \item \mbox{$P\to S$ transitions:} The transitions of total spin varying from $P$-wave to $S$-wave are small, only 0.03 keV, but the total spin invariant processes are 5 to 6 times that of them.
    
    \item \mbox{$X\to P$ transitions:} For decays to $P$-wave final states
    \begin{itemize}
        \item The transition $\Omega_{bbb}(15093)5/2^- \to \Omega_{bbb}(14687)1/2^-\gamma$ is enhanced due to the $F$-wave nature of the initial state.
        \item Negative-parity initial states show stronger transitions to $\Omega_{bbb}(14692)3/2^-\gamma$.
    \end{itemize}
    
    \item \mbox{$D\to 2S$ transitions:} Higher excited $D$-wave to $2S$-wave transitions remain significant.
\end{itemize}
Transitions to higher excited states exhibit systematically smaller widths, as shown in the table. However, there remain order-of-magnitude discrepancies between our results and those of Ref.~\cite{Liu:2019vtx}, which can be attributed to similar reasons as in the case of $\Omega_{ccc}$.

    \begin{table*}
		\caption{Radiative decay widths (in keV) for $\Omega_{ccc}$ baryons with masses below the $\varXi_{cc}^{++}D^0$ threshold. Residual radiative decay widths for $\Omega_{ccc}$ baryons not tabulated are all below 0.005 keV. \label{tab:Radiative decay widths of ccc}}
		\begin{tabular}{c@{}|@{}c}
			\hline\hline
		\begin{tabular}{ccrcccrc}
			&&\multicolumn{2}{c}{\underline{$\Omega_{ccc}(4793)\ 3/2^+\gamma$}} &\multicolumn{2}{c}{\underline{$\Omega_{ccc}(5121)\ 1/2^-\gamma$}} &\multicolumn{2}{c}{\underline{$\Omega_{ccc}(5136)\ 3/2^-\gamma$}}\\
			$(L_t,S)$&Initial state&Our&Ref.~\cite{Liu:2019vtx}&Our&Ref.~\cite{Liu:2019vtx}&Our&Ref.~\cite{Liu:2019vtx}\\
			\hline
			$(0,\frac{1}{2})$&$\Omega_{ccc}(5354)\ 1/2^+$&4.99&-&0.18&20.14&0.50&27.43\\[0.5pt]
			$(0,\frac{3}{2})$&$\Omega_{ccc}(5269)\ 3/2^+$&0.68&-&0.02&0.002&0.03&0.010\\
			\cline{2-8}
			\multirow{2}{*}{$(2,\frac{1}{2})$}
			&$\Omega_{ccc}(5409)\ 3/2^+$&9.96&-&0.43&106.88&1.33&33.58\\
			&$\Omega_{ccc}(5409)\ 5/2^+$&24.68&-&0.47&0.25&0.94&122.10\\
			\cline{2-8}
			\multirow{4}{*}{$(2,\frac{3}{2})$}
			&$\Omega_{ccc}(5370)\ 1/2^+$&176.94&-&0.14&$<$0.001&0.07&0.04\\
			&$\Omega_{ccc}(5372)\ 3/2^+$&162.60&-&0.58&0.38&0.02&0.02\\
			&$\Omega_{ccc}(5374)\ 5/2^+$&132.83&-&0.36&0.22&0.22&0.39\\
			&$\Omega_{ccc}(5378)\ 7/2^+$&138.47&-&0.00&$<$0.001&0.39&0.80\\
			\cline{2-8}
			\multirow{2}{*}{$(1,\frac{1}{2})$}
			&$\Omega_{ccc}(5121)\ 1/2^-$&2.52&3.10&-&-&-&-\\
			&$\Omega_{ccc}(5136)\ 3/2^-$&3.05&4.07&0.00&-&-&-\\
		\end{tabular}&\renewcommand\arraystretch{1.1}%
		\begin{tabular}{cc}
			&$\Omega_{ccc}(5269)\ 3/2^+\gamma$\\
			\hline
			$\Omega_{ccc}(5370)\ 1/2^+$&0.15\\
			$\Omega_{ccc}(5372)\ 3/2^+$&0.18\\
			$\Omega_{ccc}(5374)\ 5/2^+$&0.18\\
			$\Omega_{ccc}(5378)\ 7/2^+$&0.24\\
			$\Omega_{ccc}(5409)\ 3/2^+$&0.02\\
			$\Omega_{ccc}(5409)\ 5/2^+$&0.09\\
			\hline
			\multicolumn{2}{c}{$\Omega_{ccc}(5409)\ 3/2^+\to\Omega_{ccc}(5354)\ 1/2^+\gamma$ 0.01}\\
			\multicolumn{2}{c}{$\Omega_{ccc}(5409)\ 5/2^+\to\Omega_{ccc}(5354)\ 1/2^+\gamma$ 0.01}\\
			\multicolumn{2}{c}{$\Omega_{ccc}(5409)\ 5/2^+\to\Omega_{ccc}(5372)\ 3/2^+\gamma$ 0.01}\\
            \multicolumn{2}{c}{$\Omega_{ccc}(5409)\ 5/2^+\to\Omega_{ccc}(5378)\ 7/2^+\gamma$ 0.01}\\
		\end{tabular}\\
		\hline\hline
		\end{tabular}
    \end{table*}
	
	\begin{table*}
		\caption{The radiative decay widths for $\Omega_{bbb}$ baryons with masses below 15.25 GeV (in keV). Residual radiative decay widths for $\Omega_{bbb}$ baryons not tabulated are all below 0.005 keV. \label{tab:Radiative decay widths of bbb}}
		\begin{tabular}{c@{}|@{}c@{}|@{}c}
			\hline\hline
			\begin{tabular}[t]{ccrc@{}}
				&&\multicolumn{2}{c}{\underline{$\Omega_{bbb}(14365)\ 3/2^+\gamma$}}\\
				$(L_t,S)$&Initial state&Our&Ref.~\cite{Liu:2019vtx}\\
				\hline
				\multirow{2}{*}{$(0,\frac{1}{2})$}
				&$\Omega_{bbb}(14882)\ 1/2^+$&0.01&-\\
				&$\Omega_{bbb}(15170)\ 1/2^+$&0.01&-\\
				\cline{2-4}
				\multirow{4}{*}{$(2,\frac{1}{2})$}
				&$\Omega_{bbb}(14937)\ 3/2^+$&0.01&-\\
				&$\Omega_{bbb}(14937)\ 5/2^+$&0.03&-\\
				&$\Omega_{bbb}(15209)\ 3/2^+$&0.01&-\\
				&$\Omega_{bbb}(15211)\ 5/2^+$&0.03&-\\
				\cline{2-4}
				\multirow{8}{*}{$(2,\frac{3}{2})$}
				&$\Omega_{bbb}(14898)\ 1/2^+$&4.63&-\\
				&$\Omega_{bbb}(14901)\ 3/2^+$&4.56&-\\
				&$\Omega_{bbb}(14903)\ 5/2^+$&4.45&-\\
				&$\Omega_{bbb}(14905)\ 7/2^+$&4.37&-\\
				&$\Omega_{bbb}(15190)\ 1/2^+$&2.81&-\\
				&$\Omega_{bbb}(15192)\ 3/2^+$&2.84&-\\
				&$\Omega_{bbb}(15194)\ 5/2^+$&2.84&-\\
				&$\Omega_{bbb}(15196)\ 7/2^+$&2.87&-\\
				\cline{2-4}
				\multirow{4}{*}{$(1,\frac{1}{2})$}
				&$\Omega_{bbb}(14687)\ 1/2^-$&0.02&0.035\\
				&$\Omega_{bbb}(14692)\ 3/2^-$&0.03&0.038\\
				&$\Omega_{bbb}(15021)\ 1/2^-$&0.03&-\\
				&$\Omega_{bbb}(15025)\ 3/2^-$&0.03&-\\
				\cline{2-4}
				\multirow{3}{*}{$(1,\frac{3}{2})$}
				&$\Omega_{bbb}(15041)\ 1/2^-$&0.14&-\\
				&$\Omega_{bbb}(15041)\ 3/2^-$&0.13&-\\
				&$\Omega_{bbb}(15042)\ 5/2^-$&0.12&-\\
				\cline{2-4}
				\multirow{2}{*}{$(3,\frac{3}{2})$}
				&$\Omega_{bbb}(15123)\ 3/2^-$&0.36&-\\
				&$\Omega_{bbb}(15124)\ 5/2^-$&0.35&-\\
			\end{tabular}&\renewcommand\arraystretch{1.13}%
			\begin{tabular}[t]{crc}
				&\multicolumn{2}{c}{\underline{$\Omega_{bbb}(14687)\ 1/2^-\gamma$}}\\
				$\Omega_{bbb}(15170)\ 1/2^+$&0.02&\\
				$\Omega_{bbb}(15209)\ 3/2^+$&0.03&\\
				$\Omega_{bbb}(15211)\ 5/2^+$&0.02&\\
				$\Omega_{bbb}(15180)\ 3/2^-$&0.88&\\
				$\Omega_{bbb}(15143)\ 3/2^-$&0.68&\\
				$\Omega_{bbb}(15025)\ 3/2^-$&0.42&\\
				$\Omega_{bbb}(15180)\ 5/2^-$&0.33&\\
				$\Omega_{bbb}(15093)\ 5/2^-$&3.69&\\
				\hline
				&\multicolumn{2}{c}{\underline{$\Omega_{bbb}(14692)\ 3/2^-\gamma$}}\\
				&Our&Ref.~\cite{Liu:2019vtx}\\
				\hline
				$\Omega_{bbb}(15170)\ 1/2^+$&0.02&-\\
				$\Omega_{bbb}(14937)\ 3/2^+$&0.01&2.75\\
				$\Omega_{bbb}(15209)\ 3/2^+$&0.03&-\\
				$\Omega_{bbb}(15211)\ 5/2^+$&0.09&-\\
				$\Omega_{bbb}(15143)\ 1/2^-$&1.43&-\\
				$\Omega_{bbb}(15021)\ 1/2^-$&0.76&-\\
				$\Omega_{bbb}(15180)\ 3/2^-$&0.98&-\\
				$\Omega_{bbb}(15143)\ 3/2^-$&0.64&-\\
				$\Omega_{bbb}(15025)\ 3/2^-$&0.40&-\\
				$\Omega_{bbb}(15180)\ 5/2^-$&1.37&-\\
				$\Omega_{bbb}(15093)\ 5/2^-$&1.00&-\\
			\end{tabular}&\renewcommand\arraystretch{1.13}%
			\begin{tabular}[t]{cc}
				&\underline{$\Omega_{bbb}(14792)\ 3/2^+\gamma$}\\
                $\Omega_{bbb}(14898)\ 1/2^+$&0.01\\
                $\Omega_{bbb}(14901)\ 3/2^+$&0.01\\
                $\Omega_{bbb}(14903)\ 5/2^+$&0.01\\
                $\Omega_{bbb}(14905)\ 7/2^+$&0.01\\
                $\Omega_{bbb}(15190)\ 1/2^+$&2.49\\
                $\Omega_{bbb}(15192)\ 3/2^+$&2.44\\
                $\Omega_{bbb}(15194)\ 5/2^+$&2.35\\
                $\Omega_{bbb}(15196)\ 7/2^+$&2.27\\
                $\Omega_{bbb}(15211)\ 5/2^+$&0.01\\
                $\Omega_{bbb}(15021)\ 1/2^-$&0.01\\
                $\Omega_{bbb}(15025)\ 3/2^-$&0.01\\
                $\Omega_{bbb}(15123)\ 3/2^-$&0.02\\
                $\Omega_{bbb}(15124)\ 5/2^-$&0.02\\
				\hline
				&\underline{$\Omega_{bbb}(14882)\ 1/2^+\gamma$}\\
                $\Omega_{bbb}(15143)\ 1/2^-$&0.01\\
                $\Omega_{bbb}(15209)\ 3/2^+$&1.02\\
                $\Omega_{bbb}(15211)\ 5/2^+$&1.02\\
				\hline
				&\underline{$\Omega_{bbb}(14898)\ 1/2^+\gamma$}\\
                $\Omega_{bbb}(15117)\ 3/2^+$&0.05\\
                $\Omega_{bbb}(15192)\ 3/2^+$&0.20\\
                $\Omega_{bbb}(15194)\ 5/2^+$&0.06\\
			\end{tabular}\\
			\hline
			\renewcommand\arraystretch{1.15}
			\begin{tabular}{cc@{}}
				&\underline{$\Omega_{bbb}(14901)\ 3/2^+\gamma$}\\
                $\Omega_{bbb}(15117)\ 3/2^+$&0.10\\
                $\Omega_{bbb}(15190)\ 1/2^+$&0.41\\
                $\Omega_{bbb}(15194)\ 5/2^+$&0.20\\
                $\Omega_{bbb}(15196)\ 7/2^+$&0.03\\
			\end{tabular}&
			\begin{tabular}{cc}
				&\underline{$\Omega_{bbb}(14903)\ 5/2^+\gamma$}\\
				$\Omega_{bbb}(15117)\ 3/2^+$&0.15\\
                $\Omega_{bbb}(15190)\ 1/2^+$&0.18\\
                $\Omega_{bbb}(15192)\ 3/2^+$&0.32\\
                $\Omega_{bbb}(15194)\ 5/2^+$&0.06\\
                $\Omega_{bbb}(15196)\ 7/2^+$&0.19\\
			\end{tabular}&\renewcommand\arraystretch{1.15}%
			\begin{tabular}{cc}
				&\underline{$\Omega_{bbb}(14905)\ 7/2^+\gamma$}\\
				$\Omega_{bbb}(15117)\ 3/2^+$&0.20\\
                $\Omega_{bbb}(15192)\ 3/2^+$&0.07\\
                $\Omega_{bbb}(15194)\ 5/2^+$&0.27\\
                $\Omega_{bbb}(15196)\ 7/2^+$&0.37\\
			\end{tabular}\\
			\hline
			\begin{tabular}{cc@{}}
				&\underline{$\Omega_{bbb}(14937)\ 3/2^+\gamma$}\\
                $\Omega_{bbb}(15170)\ 1/2^+$&0.22\\
                $\Omega_{bbb}(15209)\ 3/2^+$&0.27\\
                $\Omega_{bbb}(15211)\ 5/2^+$&0.08\\
			\end{tabular}&
			\begin{tabular}{cc@{}}
                &\underline{$\Omega_{bbb}(14937)\ 5/2^+\gamma$}\\
                $\Omega_{bbb}(15170)\ 1/2^+$&0.34\\
                $\Omega_{bbb}(15209)\ 3/2^+$&0.12\\
                $\Omega_{bbb}(15211)\ 5/2^+$&0.32\\
			\end{tabular}&\renewcommand\arraystretch{1.2}%
                \begin{tabular}{cc@{}}
                &\underline{$\Omega_{bbb}(15117)\ 3/2^+\gamma$}\\
                $\Omega_{bbb}(15194)\ 5/2^+$&0.01\\
                $\Omega_{bbb}(15196)\ 7/2^+$&0.01\\
			\end{tabular}\\
			\hline\hline
		\end{tabular}
	\end{table*}
    
    \subsubsection{The $\Omega_{bcc}$ baryons}
    The radiative decay widths of $\Omega_{bcc}$ baryons below the 8.58 GeV are presented in Table~\ref{tab:Radiative decay widths of bcc and bbc}. Unlike the $\Omega_{ccc}$ and $\Omega_{bbb}$ systems composed of three identical quarks, $\Omega_{bcc}$ baryons exhibit distinct symmetry properties that lead to characteristic radiative decay patterns. Both $\frac{1}{2}^+$ and $\frac{3}{2}^+$ channels contain low-lying states ($\Omega_{bcc}(8017)1/2^+$ and $\Omega_{bcc}(8030)3/2^+$) that serve as dominant final states.

    The radiative decays of the $\Omega_{bcc}$ baryons show several notable features:

\begin{itemize}
    \item \mbox{$S\to S$ transitions:} The $S\to S$ transitions for the $\Omega_{bcc}$ baryons are also weak. 
    
    \item \mbox{$D\to S$ transitions:} The processes from $D$-wave to $S$-wave yield widths of $\mathcal{O}(10)$ keV. The relative widths to $1/2^+$ versus $3/2^+$ $S$-wave final states depend on total spin of initial and final states. Four $J^P=3/2^+$ or $5/2^+$ $D$-wave states show significant $S=1/2$-$3/2$ mixing, exhibiting comparable widths to both $S$-wave final states. 
    
    \item \mbox{$P\to S$ transitions:} Five $(0,1,1)$-mode excitations decaying to two $S$-wave states produce widths of $\mathcal{O}(100)$ keV, but two $(1,0,1)$-mode states show suppressed widths ($\sim$few keV).
    \item \mbox{$S\to P$ transitions:} The $S$-wave state decays to five $P$-wave states with widths ranging from several to tens of keV.
    \item \mbox{$D\to P$ transitions:} The $D\to P$ transitions follow a clear angular momentum hierarchy:
    \begin{itemize}
        \item $|J_i-J_f|=1$: largest widths (diagonal elements)
        \item $|J_i-J_f|>1$: negligible (left of diagonal)
        \item $|J_i-J_f|<1$: intermediate (right of diagonal)
    \end{itemize}
\end{itemize}
The results above adequately demonstrate the radiative decay characteristic of two-identical-quark baryon, which is far different from three-identical-quark baryon. 
    
    \subsubsection{The $\Omega_{bbc}$ baryons}
    The radiative decay widths of the $\Omega_{bbc}$ baryons below and 11.72 GeV are shown in Table \ref{tab:Radiative decay widths of bcc and bbc}. Compared to the $\Omega_{bcc}$ baryons, the radiative decay widths of the $\Omega_{bbc}$ baryons is generally smaller, and the processes with $1S$ state as the final state have a larger partial width. This may be due to the smaller charge and larger mass of $b$ quark. 

    The radiative decays of the $\Omega_{bbc}$ baryons reveal:
\begin{itemize}
    \item \mbox{$S,D\to S$ transitions:} The radiative decay widths are generally small, typically only a few tenths of keV.

    \item \mbox{$P\to S$ transitions:} Similar to the $\Omega_{bcc}$ baryons, the widths of the five $(0,1,1)$-mode excitations are $\mathcal{O}(100)$ keV, whereas the widths of the two $(1,0,1)$-mode excitations are only a few tenths of keV.
\end{itemize}
Due to identical particle symmetry, the radiative decay behavior of $\Omega_{bbc}$ closely resembles that of $\Omega_{bcc}$. However, differences in the energy level structure—caused by the constituent quark mass (i.e., ``one heavy + two light" versus ``two heavy + one light")—may introduce slight variations in their radiative decay behaviors.
    
    \begin{table*}
		\caption{The radiative decay widths of the $\Omega_{bcc}$ baryons below 8.58 GeV and the $\Omega_{bbc}$ baryons below 11.72 GeV in units of keV. All unlisted decay widths for these states are below 0.05 keV. \label{tab:Radiative decay widths of bcc and bbc}}
		\begin{tabular}{c@{}|@{}c}
			\hline\hline\renewcommand\arraystretch{1.04}
			\begin{tabular}[t]{cccc}
				$(L_t,S)$&Initial state&$\Omega_{bcc}(8017)\ 1/2^+\gamma$&$\Omega_{bcc}(8030)\ 3/2^+\gamma$\\
				\hline
				\multirow{2}{*}{$(0,\frac{1}{2})$}
				&$\Omega_{bcc}(8463)\ 1/2^+ $&0.2&0.0\\
				&$\Omega_{bcc}(8564)\ 1/2^+ $&0.8&1.1\\
				\cline{2-4}
				$(0,\frac{3}{2})$&$\Omega_{bcc}(8469)\ 3/2^+ $&0.2&0.2\\
				\cline{2-4}
				\multirow{2}{*}{$(2,\frac{1}{2})$}
				&$\Omega_{bcc}(8549)\ 3/2^+ $&22.9&15.0\\
				&$\Omega_{bcc}(8548)\ 5/2^+ $&19.1&6.0\\
				\cline{2-4}
				\multirow{4}{*}{$(2,\frac{3}{2})$}
				&$\Omega_{bcc}(8546)\ 1/2^+ $&0.0&31.0\\
				&$\Omega_{bcc}(8545)\ 3/2^+ $&12.7&15.0\\
				&$\Omega_{bcc}(8550)\ 5/2^+ $&4.7&22.6\\
				&$\Omega_{bcc}(8552)\ 7/2^+ $&0.0&27.7\\
				\cline{2-4}
				\multirow{2}{*}{$(1,\frac{1}{2})$}
				&$\Omega_{bcc}(8319)\ 1/2^- $&141.5&37.2\\
				&$\Omega_{bcc}(8322)\ 3/2^- $&135.9&24.2\\
				\cline{2-4}
				\multirow{3}{*}{$(1,\frac{3}{2})$}
				&$\Omega_{bcc}(8310)\ 1/2^- $&32.4&122.6\\
				&$\Omega_{bcc}(8323)\ 3/2^- $&19.3&134.8\\
				&$\Omega_{bcc}(8330)\ 5/2^- $&0.1&150.9\\
				\cline{2-4}
				\multirow{2}{*}{$(1,\frac{1}{2})$}
				&$\Omega_{bcc}(8389)\ 1/2^- $&1.9&1.9\\
				&$\Omega_{bcc}(8397)\ 3/2^- $&2.9&6.1\\
				\hline
				\multicolumn{3}{l}{$\Omega_{bcc}(8545)\ 3/2^+ \to\Omega_{bcc}(8389)\ 1/2^-\gamma$}&0.5\\
				\multicolumn{3}{l}{$\Omega_{bcc}(8546)\ 1/2^+ \to\Omega_{bcc}(8389)\ 1/2^-\gamma$}&0.2\\
				\multicolumn{3}{l}{$\Omega_{bcc}(8564)\ 1/2^+ \to\Omega_{bcc}(8389)\ 1/2^-\gamma$}&13.6\\
				\multicolumn{3}{l}{$\Omega_{bcc}(8546)\ 1/2^+ \to\Omega_{bcc}(8397)\ 3/2^-\gamma$}&0.2\\
				\multicolumn{3}{l}{$\Omega_{bcc}(8550)\ 5/2^+ \to\Omega_{bcc}(8397)\ 3/2^-\gamma$}&0.1\\
				\multicolumn{3}{l}{$\Omega_{bcc}(8564)\ 1/2^+ \to\Omega_{bcc}(8397)\ 3/2^-\gamma$}&25.6\\
				\hline
				\multicolumn{3}{l}{$\Omega_{bbc}(11700)\ 3/2^+ \to\Omega_{bbc}(11496)\ 1/2^-\gamma$}&0.1\\
			\end{tabular}&
			\begin{tabular}[t]{cccc}
				$(L_t,S)$&Initial state&$\Omega_{bbc}(11204)\ 1/2^+\gamma$&$\Omega_{bbc}(11221)\ 3/2^+\gamma$\\
				\hline
				$(0,\frac{3}{2})$&$\Omega_{bbc}(11627)\ 3/2^+ $&0.1&0.0\\
				\cline{2-4}
				\multirow{2}{*}{$(2,\frac{1}{2})$}
				&$\Omega_{bbc}(11700)\ 3/2^+ $&0.7&0.4\\
				&$\Omega_{bbc}(11699)\ 5/2^+ $&0.9&0.1\\
				\cline{2-4}
				\multirow{4}{*}{$(2,\frac{3}{2})$}
				&$\Omega_{bbc}(11702)\ 1/2^+ $&0.0&1.2\\
				&$\Omega_{bbc}(11708)\ 3/2^+ $&0.3&0.7\\
				&$\Omega_{bbc}(11712)\ 5/2^+ $&0.1&0.7\\
				&$\Omega_{bbc}(11715)\ 7/2^+ $&0.0&0.8\\
				\cline{2-4}
				\multirow{2}{*}{$(1,\frac{1}{2})$}
				&$\Omega_{bbc}(11571)\ 1/2^- $&131.7&12.8\\
				&$\Omega_{bbc}(11578)\ 3/2^- $&126.3&65.0\\
				\cline{2-4}
				\multirow{3}{*}{$(1,\frac{3}{2})$}
				&$\Omega_{bbc}(11559)\ 1/2^- $&16.7&147.6\\
				&$\Omega_{bbc}(11576)\ 3/2^- $&45.7&98.5\\
				&$\Omega_{bbc}(11584)\ 5/2^- $&0.6&145.3\\
				\cline{2-4}
				\multirow{2}{*}{$(1,\frac{1}{2})$}
				&$\Omega_{bbc}(11496)\ 1/2^- $&0.3&0.4\\
				&$\Omega_{bbc}(11506)\ 3/2^- $&0.1&0.0\\
				\hline
				
				\multicolumn{3}{l}{$\Omega_{bbc}(11627)\ 3/2^+ \to\Omega_{bbc}(11559)\ 1/2^-\gamma$}&0.1\\
				\multicolumn{3}{l}{$\Omega_{bbc}(11621)\ 1/2^+ \to\Omega_{bbc}(11571)\ 1/2^-\gamma$}&0.1\\
				\multicolumn{3}{l}{$\Omega_{bbc}(11700)\ 3/2^+ \to\Omega_{bbc}(11571)\ 1/2^-\gamma$}&0.1\\
				\multicolumn{3}{l}{$\Omega_{bbc}(11708)\ 3/2^+ \to\Omega_{bbc}(11571)\ 1/2^-\gamma$}&0.1\\
				\multicolumn{3}{l}{$\Omega_{bbc}(11708)\ 3/2^+ \to\Omega_{bbc}(11576)\ 3/2^-\gamma$}&0.1\\
				\multicolumn{3}{l}{$\Omega_{bbc}(11712)\ 5/2^+ \to\Omega_{bbc}(11576)\ 3/2^-\gamma$}&0.2\\
				\multicolumn{3}{l}{$\Omega_{bbc}(11621)\ 1/2^+ \to\Omega_{bbc}(11578)\ 3/2^-\gamma$}&0.1\\
				\multicolumn{3}{l}{$\Omega_{bbc}(11699)\ 5/2^+ \to\Omega_{bbc}(11578)\ 3/2^-\gamma$}&0.1\\
				\multicolumn{3}{l}{$\Omega_{bbc}(11627)\ 3/2^+ \to\Omega_{bbc}(11584)\ 5/2^-\gamma$}&0.1\\
				\multicolumn{3}{l}{$\Omega_{bbc}(11715)\ 7/2^+ \to\Omega_{bbc}(11584)\ 5/2^-\gamma$}&0.2\\
			\end{tabular}\\
			\hline\hline
			\multicolumn{2}{l}{\begin{tabular}{ccccccc}
					$(L_t,S)$&Initial state&$\Omega_{bcc}(8310)\ 1/2^-\gamma$&$\Omega_{bcc}(8319)\ 1/2^-\gamma$&$\Omega_{bcc}(8322)\ 3/2^-\gamma$&$\Omega_{bcc}(8323)\ 3/2^-\gamma$&$\Omega_{bcc}(8330)\ 5/2^-\gamma$\\
					\hline
					\multirow{3}{*}{$(0,\frac{1}{2}/\frac{3}{2})$}
					&$\Omega_{bcc}(8463)\ 1/2^+ $&5.0&16.0&36.8&6.4&0.0\\
					&$\Omega_{bcc}(8564)\ 1/2^+ $&0.2&0.5&0.7&1.2&0.1\\
					&$\Omega_{bcc}(8469)\ 3/2^+ $&9.6&2.1&2.8&18.6&32.1\\
					\cline{2-7}
					\multirow{6}{*}{$(2,\frac{1}{2}/\frac{3}{2})$}
					&$\Omega_{bcc}(8546)\ 1/2^+ $&146.3&32.3&4.4&29.8&0.0\\
					&$\Omega_{bcc}(8545)\ 3/2^+ $&98.1&19.3&0.8&68.3&5.4\\
					&$\Omega_{bcc}(8549)\ 3/2^+ $&1.1&129.3&45.9&23.8&5.3\\
					&$\Omega_{bcc}(8548)\ 5/2^+ $&0.0&0.2&170.4&0.0&11.5\\
					&$\Omega_{bcc}(8550)\ 5/2^+ $&0.1&0.0&2.4&133.3&44.2\\
					&$\Omega_{bcc}(8552)\ 7/2^+ $&0.0&0.0&0.1&0.0&172.1\\
			\end{tabular}}\\
			\hline\hline
		\end{tabular}
	\end{table*}
    \section{SUMMARY}
    Within the framework of the nonrelativistic quark model, we systematically investigate the mass spectra of the triply heavy baryons up to $D$-wave states using the Gaussian expansion method. Subsequently, the radiative decay widths for states up to $1D$ are calculated utilizing the wave functions obtained from the mass spectrum computations.

Key findings regarding the spectroscopic properties, as elaborated in preceding sections, are summarized as follows:
\begin{itemize}
    \item \textbf{$\Omega_{ccc}$ and $\Omega_{bbb}$ baryons:}
    \begin{enumerate}
        \item The predicted masses for low-lying states agree with lattice QCD results, while those for excited states are generally lower—a trend consistent with most potential model predictions.
        \item We establish that $J^P=1/2^+$ baryons composed of three identical quarks still possess $S$-wave bound states, correcting a misinterpretation in earlier literature.
    \end{enumerate}
    
    \item \textbf{$\Omega_{bcc}$ and $\Omega_{bbc}$ baryons:}
    \begin{enumerate}
        \item Our calculated mass spectra are consistent with lattice results within their respective uncertainties.
        \item We demonstrate the existence of all excitation modes for baryons with two identical quarks, with certain modes exhibiting mixing effects. This contrasts with approaches in other works that explicitly exclude specific modes.
    \end{enumerate}
\end{itemize}
The predicted spectra await experimental verification, particularly at future heavy-quark production facilities.

For radiative decays, the principal conclusions are:
\begin{enumerate}
    \item Radiative decay patterns differ fundamentally between baryons with three identical quarks and those with two identical quarks, governed by distinct underlying mechanisms.
    \item Our calculated radiative decay widths for $\Omega_{ccc}$ and $\Omega_{bbb}$ baryons exhibit discrepancies with Ref. \cite{Liu:2019vtx}, likely attributable to its omission of identical particle symmetry considerations.
    \item Mixing between the total spin states $S=1/2$ and $S=3/2$ plays a crucial role in electric dipole transitions. Consequently, accounting for this mixing is essential for accurate radiative decay width calculations.
\end{enumerate}
We anticipate that our radiative decay predictions will provide valuable guidance for future experiments, enabling verification of decay behaviors governed by distinct symmetry principles. 

\begin{acknowledgments}
This work is supported by the National Natural Science Foundation of China under Grant Nos. 12335001, 12247101, and 12405098,  the ‘111 Center’ under Grant No. B20063, the Natural Science Foundation of Gansu Province (No. 22JR5RA389, No. 25JRRA799), the Talent Scientific Fund of Lanzhou University, the fundamental Research Funds for the Central Universities (No. lzujbky-2023-stlt01), and the project for top-notch innovative talents of Gansu province.
\end{acknowledgments}
    
    \bibliography{References}
    
\end{document}